\documentclass[twocolumn,aps,prb,floatfix]{revtex4-2}
\usepackage{newtxtext}
\usepackage{amsmath,amsfonts,amssymb}
\usepackage{mathrsfs}
\usepackage{soul,bm,array,graphicx,bbold,multirow}
\usepackage{bbm}
\usepackage[normalem]{ulem}
\usepackage[makeroom]{cancel}
\usepackage[usenames,dvipsnames]{xcolor}
\usepackage[colorlinks=true,citecolor=blue,linkcolor=red]{hyperref}
\usepackage{pdfpages}
\usepackage{lipsum}
\usepackage{tikz}
\usepackage{booktabs}
\usepackage{enumitem}
\usetikzlibrary{tikzmark, calc, decorations.pathreplacing}
\makeatletter
\AtBeginDocument{\let\LS@rot\@undefined}
\makeatother

\newcolumntype{x}[1]{>{\centering\let\newline\\\arraybackslash\hspace{0pt}}p{#1}}
\DeclareMathAlphabet{\mathbbold}{U}{bbold}{m}{n}
\newcounter{subeqn} %
\renewcommand{\Re}{\operatorname{Re}}
\renewcommand{\Im}{\operatorname{Im}} 
\makeatletter
\@addtoreset{subeqn}{equation}
\makeatother

\makeatletter
\def\l@subsection#1#2{}
\def\l@subsubsection#1#2{}
\def\l@f@section{%
  \addpenalty{\@secpenalty}%
  \addvspace{0.25em plus\p@}%
}
\makeatother

\begin{document}

\title{Universal Generalized Brillouin Zone Theory I: Review of the Spectral Approach}

\author{Zeqi Xu$^1$}
\author{Jiangping Hu$^{2,3}$}
\email[Corresponding author: ]{jphu@iphy.ac.cn}
\author{Zhesen Yang$^{1,4}$}
\email[Corresponding author: ]{yangzs@xmu.edu.cn}

\affiliation{$^1$Department of Physics, Xiamen University, Xiamen 361005, Fujian Province, China}
\affiliation{$^2$Beijing National Laboratory for Condensed Matter Physics and Institute of Physics, Chinese Academy of Sciences, Beijing 100190, China}
\affiliation{$^3$New Cornerstone Science Laboratory, Beijing, 100190, China}
\affiliation{$^4$Asia Pacific Center for Theoretical Physics, Pohang 37673, Korea}

\date{\today}

\begin{abstract}
This series of papers aims to establish a universal generalized Brillouin zone (GBZ) theory for higher-dimensional non-Hermitian systems.
As the starting point of this series, we emphasize a fundamental question: while the conventional one-dimensional (1D) GBZ condition, $|\beta_p| = |\beta_{p+1}|$, is well known to fail in two dimensions, how does this breakdown actually occur as a system gradually crosses over from 1D to 2D?
Investigating this question reveals that the existing 1D GBZ theory itself remains incomplete.
In this first paper, we therefore systematically review the 1D spectral approach and clarify where the underlying difficulties lie.
Our work identifies the key challenges that motivate the wavefunction approach developed in Paper~II.
\end{abstract}

\maketitle
%{\hypersetup{linkcolor=black}%
%\tableofcontents}

\section{Introduction}\label{sec:Introduction}

The energy spectrum and wavefunctions provide the starting point for understanding the physical properties of a lattice model.
For a finite system with open boundary conditions (OBCs), they can be obtained by diagonalizing its OBC Hamiltonian $H_{\mathrm{OBC}}$.
With $L$ unit cells and $n$ internal degrees of freedom per cell, this is an $nL\times nL$ matrix.
Therefore, direct diagonalization becomes impractical for macroscopically large systems, making analytical methods essential.

For periodic Hermitian lattices, Bloch's theorem makes this problem tractable.
Under periodic boundary conditions (PBCs), one need only diagonalize the $n\times n$ Bloch Hamiltonian $H(\bm{k})$ at each crystal momentum $\bm{k}$.
This simplification lies at the heart of energy band theory and much of solid-state physics~\cite{bloch1929quantenmechanik,kittel1976introduction,ashcroft1976solid}.

To apply this description to an open system, one assumes that changing the boundary conditions from PBC to OBC leaves the bulk spectrum unchanged in the thermodynamic limit~\cite{ashcroft1976solid}.
Although this is generally true for Hermitian systems, it can fail in non-Hermitian systems.
A prominent example is the non-Hermitian skin effect (NHSE), in which a macroscopic number of eigenstates become exponentially localized near the boundaries~\cite{ZhongWang2018PRL,Shunyu2018PRL,Gong2018PRX,Kunst2018PRL,Martinez2018PRB,Deng2019PRB,Longhi2019PRR,Kawabata2019PRX,Fei2019PRL,Murakami2019PRL,SongFei2019PRL,ChingHua2019PRB,Kai2020PRL,Zhesen2020PRL,Kawabata2020PRBa,Yifei2020PRL,Okuma2020PRL,Borgnia2020PRL,Zirnstein2021PRL,Liu2024PRL,Review1,Review2,Review3,Review4}. 
In such systems, the PBC and OBC spectra can differ substantially in the complex energy plane.

Therefore, the NHSE calls for an extension of the Bloch description.
To accommodate exponential localization, the Brillouin zone (BZ) is extended from real to complex momenta~\cite{ZhongWang2018PRL}.
In one dimension (1D), we write the corresponding translation factor as
\begin{equation}
	\beta \equiv e^{ik}e^{\mu},
\end{equation}
where $k$ and $\mu$ are real.
The phase $k$ describes spatial oscillations, while $\mu=\ln|\beta|$ sets the rate of exponential growth or decay.

\begin{figure*}[t]
	\begin{center}
		\includegraphics[width=1\linewidth]{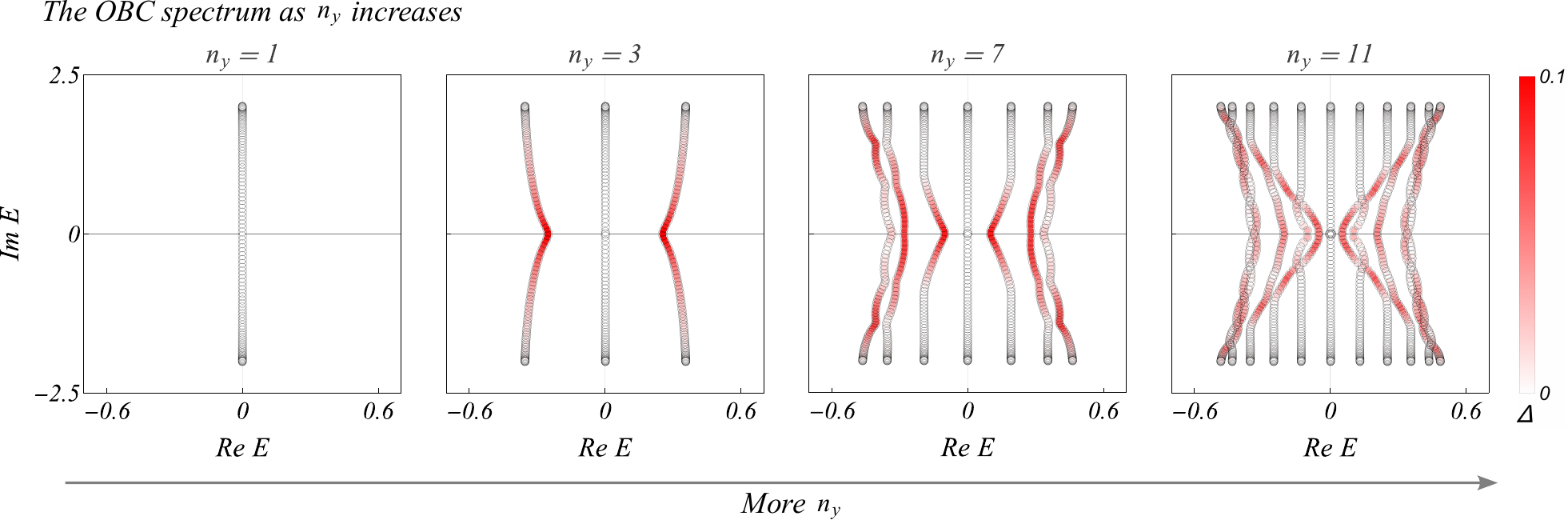}
		\par\end{center}
	\protect\caption{OBC energy spectra (black circles) for fixed $L_x=100$ and varying widths $n_y=1, 3, 7$, and $11$. Red color intensity encodes the magnitude of the 1D GBZ deviation. Parameters: $t_x=t_{-x}=i, t_y=t_{-y}=1/4$. All computations use 16-digit numerical precision.
	} 
	\label{Fig_1Dto2D}
\end{figure*}

In the conventional BZ under PBCs, we have $\mu=0$, so that $\beta$ traces the unit circle.
However, under OBCs, $\mu$ generally depends on $k$ and need not vanish.
This deformation naturally leads to the notion of the generalized Brillouin zone (GBZ), defined as the set of complex translation factors that reproduces the bulk OBC spectrum in the thermodynamic limit~\cite{ZhongWang2018PRL,Murakami2019PRL,Zhesen2020PRL}.
If the GBZ radius is a single-valued function of $k$, the GBZ of the $m$-th band can be written as
\begin{equation}
	\beta_{m,\rm GBZ}(k) \equiv e^{ik}e^{\mu_{m,\rm GBZ}(k)},
\end{equation}
where $m=1,\ldots,n$ labels the bands.
The bulk OBC spectrum is then obtained by evaluating the corresponding dispersion relation along each GBZ:
\begin{equation}
	\sigma_{\mathrm{OBC,bulk}} = \bigcup_m \left\{ E_m(\beta_{m,\rm GBZ}(k)) \,\Big|\, k \in [0, 2\pi) \right\}.
\end{equation}
Here $E_m(\beta)$ denotes the dispersion branch associated with the $m$-th GBZ.
The central question is therefore: what determines $\mu_{m,\rm GBZ}(k)$?

The standard approach starts from the characteristic equation (ChE), $\det[E\mathbbm{I}_n-H(\beta)]=0$.
We order its roots by increasing modulus:
\begin{equation*}
	|\beta_1(E)|\le |\beta_2(E)|\le\cdots\le |\beta_{n_s}(E)|,
\end{equation*}
where $n_s$ is the number of roots, counted with multiplicity.
For an $n$-band model with full-rank longest-range hopping matrices, $n_s=n(p+q)$, where $p$ and $q$ are the maximum hopping ranges to the right and left, respectively.
For generic models satisfying the full-rank condition discussed in Sec.~\ref{sec:1D Open Boundary Problem and Equations}, the bulk OBC spectrum in the thermodynamic limit is characterized by the conventional GBZ condition~\cite{ZhongWang2018PRL,Murakami2019PRL,Zhesen2020PRL}:
\begin{equation}\label{1DGBZ}
	|\beta_{np}(E)|=|\beta_{np+1}(E)|.
\end{equation}
This condition describes the thermodynamic bulk spectrum and does not by itself determine finite-size corrections or the quantization of topological edge states.
Even in this limit, however, exceptions are known~\cite{Kawabata2020PRBa,Yifei2020PRL}.
Understanding when and why the conventional condition fails is a central goal of this series.

\subsection{Motivating example}

The 1D GBZ theory is well established for conventional cases~\cite{ZhongWang2018PRL,Murakami2019PRL,Zhesen2020PRL,Yifei2020PRL,Kai2020PRL,Okuma2020PRL}.
However, extending it to two dimensions (2D) and beyond remains a central problem, with several approaches addressing different aspects~\cite{Kawabata2020PRBb,Kai2022NC,Xue2022PRL,Hui2023PRL,Yokomizo2023PRB,ZhangKai2023PRL,Kawabata2023PRX,Fang2023PRB,Yuncheng2024PRB,Hu2024PRL,Hongyi2024PRX,Haiping2025SciB,Kai2025PRX,Yuncheng2024arXiv,Zeqi2023arXiv,ChangShu2024arXiv,Chenyang2025arXiv}.
This series aims to develop a unified GBZ theory for higher-dimensional systems and clarify its relation to other approaches.
A detailed review of existing 2D theories will be given in subsequent papers.

Although there is no well-accepted 2D GBZ theory, it is clear that the 1D GBZ condition in Eq.~(\ref{1DGBZ}) cannot be generalized to higher dimensions.
Therefore, our motivating question is: how does the 1D GBZ condition in Eq.~(\ref{1DGBZ}) break down as the system crosses over from 1D to 2D, and how can this breakdown be characterized?

As a concrete example, we consider a 2D single-band tight-binding model with the non-Bloch Hamiltonian
\begin{equation} 
	H(\beta_x,\beta_y)=t_x\beta_x+t_{-x}/\beta_x+t_y\beta_x\beta_y+t_{-y}/(\beta_x\beta_y).
\end{equation} 
Here $\beta_x$ and $\beta_y$ are the translation factors in the two lattice directions, and $t_{\pm x}$ and $t_{\pm y}$ are the hopping amplitudes.
We impose OBCs in both directions on a rectangular lattice with $L_x$ sites along $x$ and $n_y$ sites along $y$.
We group the $n_y$ sites at each $x$ coordinate into one supercell, reducing the problem to an effective 1D model.
Increasing $n_y$ can be regarded as a crossover from 1D to 2D.

Therefore, replacing translation along $x$ by the complex factor $\beta_x$, the transverse chain has the hopping amplitudes
\begin{equation}\label{Tycoeffs}
\begin{aligned}
	T_{y,0}(\beta_x) &= t_x\beta_x + t_{-x}/\beta_x, \\
	T_{y,+1}(\beta_x) &= t_y\beta_x, \\
	T_{y,-1}(\beta_x) &= t_{-y}/\beta_x,
\end{aligned}
\end{equation}
where $T_{y,0}$ is the onsite energy and $T_{y,\pm1}$ multiply $(\beta_y)^{\pm1}$.
The resulting $n_y\times n_y$ Hamiltonian reads
\begin{equation}\label{Hny}
	H_{n_y}(\beta_x)=
	\begin{pmatrix}
		T_{y,0} & T_{y,+1}  \\[-0.5ex]
		T_{y,-1} & T_{y,0} & \ddots \\[-0.5ex]
		& \ddots & \ddots &\ddots \\[-1ex]
		& & \ddots & T_{y,0} & T_{y,+1} \\[0.5ex]
		& & & T_{y,-1} & T_{y,0}
	\end{pmatrix}_{n_y},
\end{equation}
where the matrix elements depend on $\beta_x$ through Eq.~(\ref{Tycoeffs}).

Fig.~\ref{Fig_1Dto2D} shows the OBC spectra for $n_y=1,3,7$, and $11$ at fixed $L_x=100$.
As $n_y$ increases, the spectrum evolves from a curve toward an area-filling pattern, indicating the crossover from 1D to 2D.
To quantify deviations from Eq.~(\ref{1DGBZ}), we substitute each finite-size eigenvalue $E$ into the ChE,
\begin{equation}
	\det[E\mathbbm{I}_{n_y}-H_{n_y}(\beta_x)]=0
\end{equation}
and obtain $2n_y$ roots $\{\beta_{x,i}(E)\}$, ordered by increasing modulus.

We define a quantity $\Delta(E)$ to characterize the deviation from the 1D GBZ condition:
\begin{equation}
	\Delta(E) =|\beta_{x,n_yp+1}(E)|-|\beta_{x,n_yp}(E)|.
\end{equation} 
Here $p=1$ for this model.
In Fig.~\ref{Fig_1Dto2D}, darker red indicates a larger deviation.
As $n_y$ increases, more OBC eigenvalues depart from the 1D GBZ condition, with a clear dependence on their position in the spectrum.
The eigenvalues in the triangular regions near the imaginary axis nearly satisfy Eq.~(\ref{1DGBZ}), whereas those in the left and right wings show larger deviations.
These deviations increase with the transverse width.

This example shows why the 1D problem must be examined before extending the theory to higher dimensions.
It raises three questions:

(i) For fixed $n_y$, does the spectrum converge to the 1D GBZ prediction as $L_x \to \infty$, or do deviations persist in this limit?

(ii) If convergence occurs, how large must $L_x$ be for the thermodynamic prediction to become reliable?

(iii) If the required size is too large for direct numerical verification, how can the theory be tested numerically?

\noindent These three questions will also be addressed in the final section.  

\subsection{Organization of the paper}

In this paper, we review the 1D spectral approach, demonstrate that the existing theory cannot resolve these questions, and clarify where the fundamental difficulties lie.

Section~\ref{sec:1D Open Boundary Problem and Equations} formulates the 1D open boundary problem.
Sections~\ref{sec:BME} and~\ref{sec:Exact Calculations} derive the boundary matrix equation and present exact finite-size calculations via Vieta's formulas.
Sections~\ref{sec:Spectral Approach}, \ref{subsec:Alternative Formulations}, and \ref{sec:Limitations and Challenges of the Spectral Approach} review the spectral approach in the thermodynamic limit, its alternative formulations, and its limitations.
Section~\ref{sec:Conclusion} summarizes the results, discusses the open questions, and outlines Papers~II--IV.

\section{1D Open Boundary Problem}\label{sec:1D Open Boundary Problem and Equations}

In the motivating example, we described a strip by grouping its $n_y$ sites at each $x$ coordinate into one supercell, reducing the problem to an effective 1D model. We now consider the general 1D open boundary problem, which is also applicable to these quasi-1D systems.

Consider a general 1D $n$-band model whose non-Bloch Hamiltonian is a matrix Laurent polynomial of the complex variable $\beta$:
\begin{equation}
	H(\beta) = \sum_{i=-p}^{q} T_i \beta^i,
\end{equation}
where $p,q\in\mathbb{N}^+$ are the maximum hopping ranges to the right and left, respectively.
The $n\times n$ matrix $T_i$ describes hopping from unit cell $x+i$ to unit cell $x$.
Here $n$ counts the internal states per unit cell, including spin, orbital, sublattice, and layer degrees of freedom.
The eigenenergy $E$ of $H(\beta)$ satisfies the ChE
\begin{equation}\label{ChE}
	f(E, \beta) := \det[E\mathbbm{I}_n - H(\beta)] = 0.
\end{equation} 
When the longest-range hopping matrices satisfy the full-rank condition
\begin{equation}\label{FRC}
	\det[T_{-p}] \neq 0, \quad \det[T_q] \neq 0,
\end{equation}
multiplying $f(E,\beta)$ by $\beta^{np}$ gives a polynomial of degree $n_s\equiv n(p+q)$ with a nonzero constant term.
Thus, for each $E$, the ChE has exactly $n_s$ finite, nonzero roots, counted with multiplicity.
Unless stated otherwise, we assume the full-rank condition throughout.
An example with rank-deficient hopping matrices is the non-Hermitian Su--Schrieffer--Heeger (SSH) model, discussed in Sec.~\ref{subsec:SSH_BME}.

The hopping matrices $\{T_i\}$ also determine the real-space Hamiltonian under OBCs.
For a chain of $L$ unit cells, with $L\geq p+q$ so that the boundary regions do not overlap, the eigenvalue equation reads
\begin{widetext}
	\begin{samepage}
		\begin{equation}\label{HOBC}
	\left[ \begin{array}{ccccccccc}
		T_0    & T_1    & \cdots & T_q    & & & & &  \\ 
		T_{-1} & T_0    & \ddots & T_{q-1}& T_q  & & & & \\
		\vdots & \ddots & \ddots & \ddots & \ddots & \ddots & & & \\[-0.2ex]  \hline \\[-3.2ex]
		T_{-p} & T_{-p+1} & \ddots & T_0 & T_1 & \ddots & T_q & &  \\
		& \ddots & \ddots & \ddots & \ddots & \ddots & \ddots & \ddots & \\
		& & T_{-p} & \ddots & T_{-1} & T_0 & \ddots & T_{q-1} & T_q  \\[0.3ex] \hline  \\[-2.5ex]
		& & & \ddots & \ddots & \ddots & \ddots & \ddots & \vdots \\
		& & & & T_{-p} & T_{-p+1} & \ddots & T_0  & T_1 \\ \\[-1.2ex]
		& & & & & T_{-p} & \cdots & T_{-1} & T_0 \\
	\end{array} \right]
	\left[ \begin{array}{c}
		\psi(1) \\ [1.5ex]
		\psi(2) \\ 
		\vdots \\
		\hline \\[-1.5ex]
		\psi(p{+}1) \\ [0.5ex]
		\vdots \\[1ex]
		\psi(L{-}q) \\[0.5ex] 
		\hline  \\[-2.5ex]
		\vdots \\ [1.2ex]
		\psi(L{-}1) \\ \\[-1.2ex]
		\psi(L) \\
	\end{array} \right]
	= E
	\left[ \begin{array}{c}
		\tikzmark{bL}\psi(1) \\ [1.5ex]
		\psi(2) \\ 
		\vdots\tikzmark{bLend} \\
		\hline \\[-1.5ex]
		\tikzmark{bB}\psi(p{+}1) \\ [0.5ex]
		\vdots \\[1ex]
		\psi(L{-}q)\tikzmark{bBend} \\[0.5ex] 
		\hline  \\[-2.5ex]
		\tikzmark{bR}\vdots \\ [1.2ex]
		\psi(L{-}1) \\ \\[-1.2ex]
		\psi(L)\tikzmark{bRend} \\
	\end{array} \right]\hspace{75pt}
	\end{equation}
\begin{tikzpicture}[overlay, remember picture]
\coordinate (Lstart) at ([xshift=42pt, yshift=6pt]pic cs:bL);
\coordinate (pLend)  at (pic cs:bLend);
\coordinate (Lend)   at ([yshift=-2pt]Lstart |- pLend);
\draw[decorate, decoration={brace, amplitude=4pt}, thick] (Lstart) -- (Lend) node[midway, right=5pt] {\small Left ($p$ rows)};

\coordinate (Bstart) at ([xshift=48pt, yshift=6pt]pic cs:bB);
\coordinate (pBend)  at (pic cs:bBend);
\coordinate (Bend)   at ([yshift=-2pt]Bstart |- pBend);
\draw[decorate, decoration={brace, amplitude=5pt}, thick] (Bstart) -- (Bend) node[midway, right=5pt] {\small Bulk ($L{-}p{-}q$ rows)};

\coordinate (Rstart) at ([xshift=34pt, yshift=6pt]pic cs:bR);
\coordinate (pRend)  at (pic cs:bRend);
\coordinate (Rend)   at ([yshift=-2pt]Rstart |- pRend);
\draw[decorate, decoration={brace, amplitude=4pt}, thick] (Rstart) -- (Rend) node[midway, right=5pt] {\small Right ($q$ rows)};
\end{tikzpicture}
\end{samepage}
\end{widetext}  
Here $\psi(x)\equiv[\psi_1(x),\psi_2(x),\dots,\psi_n(x)]^T\in\mathbb{C}^n$ is the wavefunction vector at unit cell $x\in\{1,2,\dots,L\}$.

The block rows of Eq.~(\ref{HOBC}) fall into three spatial regions:
\begin{itemize}
	\item[(i)] {\em Left boundary} (first $p$ rows): The boundary removes the hoppings from the left of the chain.
	\item[(ii)] {\em Bulk} (middle $L - p - q$ rows): Each row gives the same translation-invariant difference equation,
	\begin{equation}\label{BulkE}
	E\,\psi(x) = \sum_{i=-p}^{q} T_i\, \psi(x+i), \quad x=p+1,\dots,L-q.
    \end{equation}
	\item[(iii)] {\em Right boundary} (last $q$ rows): The boundary removes the hoppings from the right of the chain.
\end{itemize}
We emphasize two points about this OBC Hamiltonian.
\begin{figure}[tbp]
	\begin{center}
		\includegraphics[width=0.63\linewidth]{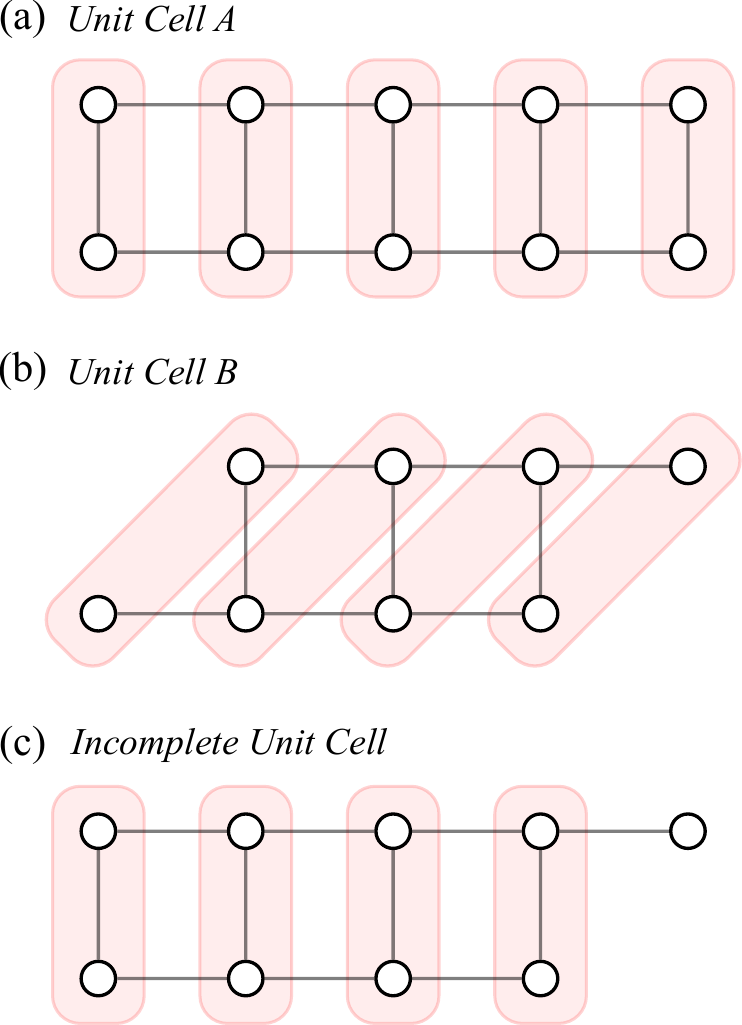}
		\par\end{center}
	\protect\caption{Unit-cell choices and OBC geometries. (a),(b) Rectangular and parallelogram groupings on a two-chain lattice. Red-shaded capsules indicate the unit cells. (c) An incomplete unit cell at the right edge, which is not described by $H_{\text{OBC}}$ in Eq.~(\ref{HOBC}).
	}
	\label{F_twoG}
\end{figure}

First, $H_{\text{OBC}}$ in Eq.~(\ref{HOBC}) describes a chain of $L$ complete unit cells.
Each cell retains all its internal degrees of freedom, including at the boundaries [Figs.~\ref{F_twoG}(a) and~(b)].

Second, for fixed $L$ and a fixed ordered internal basis, the hopping matrices $\{T_i\}$ uniquely determine $H_{\text{OBC}}$.
For a fixed microscopic lattice and hopping pattern, choosing the unit cells determines the following objects:
\begin{equation}
\begin{aligned}
	\text{Unit-cell choice} &\longrightarrow \{T_i\}\longleftrightarrow H(\beta),\\
	(\{T_i\},L) &\longrightarrow H_{\text{OBC}}(L),\\
	(\text{Unit-cell choice},L) &\longrightarrow \text{OBC geometry}.
\end{aligned}
\end{equation}
As a concrete example, consider the two-band lattice model ($n=2$)~\cite{Yifei2025PRBa,Yifei2025PRBb} in Fig.~\ref{F_twoG}.
It consists of two identical 1D chains coupled along $y$ by the hopping amplitudes $t_y$ and $t_{-y}$.
For Unit Cell~A [Fig.~\ref{F_twoG}(a)], both interchain couplings lie within the unit cell.
The non-Bloch Hamiltonian reads
\begin{equation}
	H^{A}(\beta) = \begin{bmatrix}
		h_x(\beta) & t_y\\[0.5ex]
		t_{-y} & h_x(\beta)\end{bmatrix},
\end{equation} 
where
\begin{equation}
	h_x(\beta) = t_1\beta + t_0 + t_{-1}/\beta + t_{-2}/\beta^2.
\end{equation}
Here $t_0$ is the onsite energy, while $t_1$, $t_{-1}$, and $t_{-2}$ are the intrachain hopping amplitudes.
This model has $n=2$, $p=2$ and $q=1$.
Under OBCs, this supercell chain forms a rectangular geometry.

For Unit Cell~B [Fig.~\ref{F_twoG}(b)], the $t_{\pm y}$ connect neighboring unit cells.
The Hamiltonian becomes
\begin{equation}
	H^{B}(\beta) = \begin{bmatrix}
		h_x(\beta) & t_y\beta\\[0.5ex]
		t_{-y}/\beta & h_x(\beta)\end{bmatrix},
\end{equation} 
which corresponds to a parallelogram geometry under OBCs.

\begin{figure}[t]
	\begin{center}
		\includegraphics[width=1\linewidth]{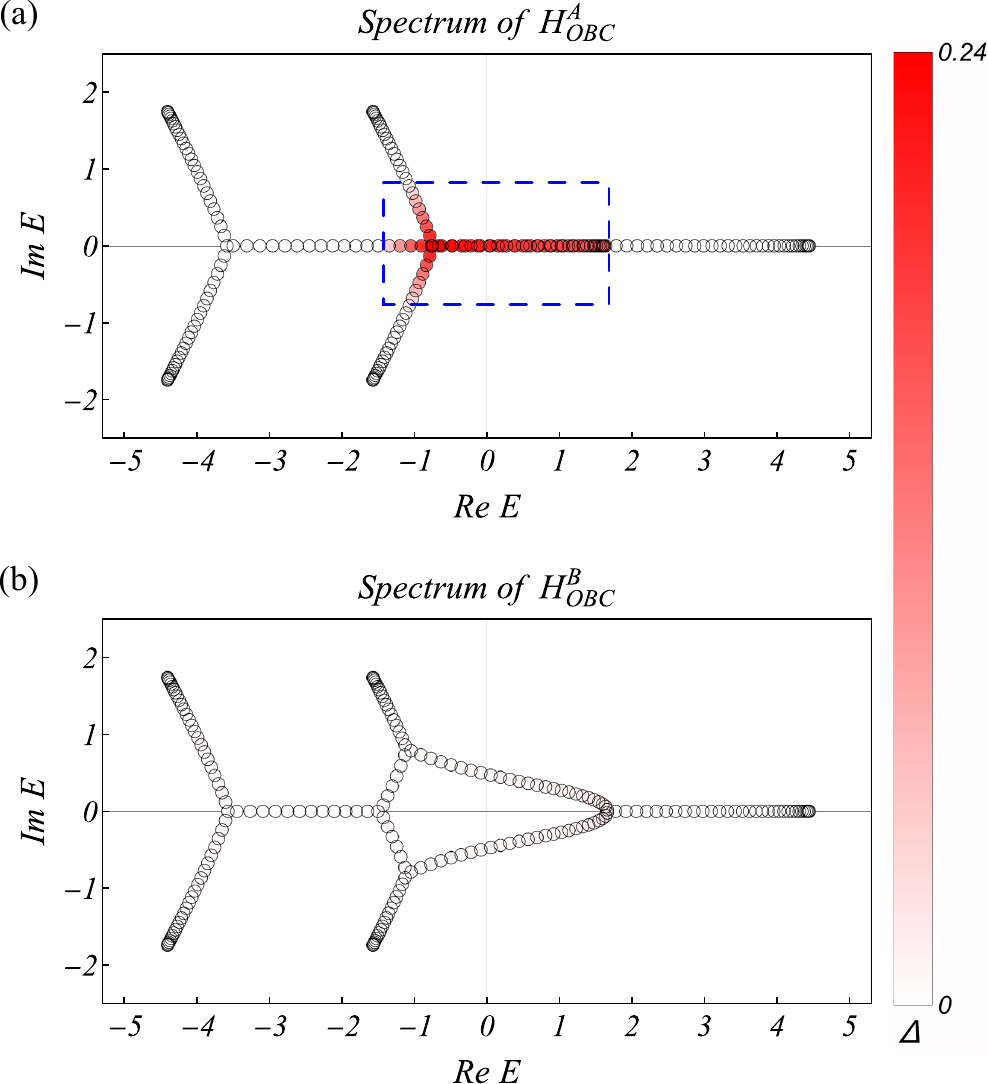}
		\par\end{center}
	\protect\caption{OBC spectra for different unit-cell choices. (a),(b) Spectra of $H_{\rm OBC}^{A}$ and $H_{\rm OBC}^{B}$ for $L=200$. Color intensity indicates the 1D GBZ deviation $\Delta$; blue dashed boxes in (a) mark intervals with nonzero $\Delta$. Parameters: $h_x(\beta)=2\beta-0.9+1/\beta+1/\beta^2$, $t_y=2$, $t_{-y}=1$. All computations use 16-digit numerical precision. 
	}
	\label{F_twoG2}
\end{figure}

The corresponding OBC spectra for $L=200$ are shown in Figs.~\ref{F_twoG2}(a) and \ref{F_twoG2}(b).
The non-Bloch Hamiltonians $H^{A}(\beta)$ and $H^{B}(\beta)$ have the same characteristic Laurent polynomial,
\begin{equation}\label{ChE_equal}
\begin{aligned}
	\det[E\mathbbm{I}_2- H^A(\beta)] &= \det[E\mathbbm{I}_2- H^B(\beta)] \\
	&= [E - h_x(\beta)]^2 - t_y t_{-y},
\end{aligned}
\end{equation}
and describe the same bulk system under PBCs.
However, their OBC Hamiltonians $H_{\text{OBC}}^{A}$ and $H_{\text{OBC}}^{B}$ nevertheless have different spectra.
For Unit Cell~A, some spectral intervals have a nonzero GBZ deviation $\Delta(E)=|\beta_5|-|\beta_4|$ at the displayed system size.
No comparable deviation is resolved for Unit Cell~B at this size.
This difference illustrates {\em unit-cell-dependent critical NHSE}.

In summary, different unit-cell choices redistribute the intracell and intercell hoppings and can produce different $\{T_i\}$, $H(\beta)$, and $H_{\text{OBC}}$.
The formulation used here assumes that each unit cell retains all $n$ internal states.
For example, Fig.~\ref{F_twoG}(c) shows an incomplete unit cell at the boundary, which is not described by $H_{\text{OBC}}$ in Eq.~(\ref{HOBC}).

\section{Boundary Matrix Equation}\label{sec:BME}
The boundary matrix equation (BME) reduces the OBC eigenvalue problem to algebraic equations for the energy and mode coefficients. 
The bulk rows of Eq.~(\ref{HOBC}) form a linear recurrence with constant coefficients.
We first solve this recurrence using generalized Bloch modes, then impose the boundary conditions to determine their superposition coefficients.
We then use the resulting boundary matrix to formulate the central questions in the thermodynamic limit.

\subsection{\texorpdfstring{Boundary Matrix for a Finite Chain}{Boundary Matrix for a Finite Chain}}

For a given energy $E$, we order the roots $\{\beta_j(E)\}_{j=1}^{n_s}$ of the ChE~(\ref{ChE}) by increasing modulus:
\begin{equation}\label{OrderBeta}
	|\beta_1| \le \dots \le |\beta_{np}| \le |\beta_{np+1}| \le \dots \le |\beta_{n_s}|.
\end{equation} 
Each root $\beta_j$ has an associated internal right eigenvector $u_j\equiv u_j^R\in\mathbb{C}^n$ satisfying
\begin{equation}\label{EigenFunction_n}
	H(\beta_j)\, u_j = E\, u_j.
\end{equation} 
We omit the superscript $R$ below because only right eigenvectors enter this construction.

Assuming that the $n_s$ roots are distinct, the OBC wavefunction can be written as
\begin{equation}\label{Psix}
	\psi(x) = \sum_{j=1}^{n_s} C_j\, (\beta_j)^x\, u_j, \quad x =1, \dots, L,
\end{equation} 
where $\{C_j\}_{j=1}^{n_s}$ are initially undetermined.
We first verify that this ansatz satisfies the translationally invariant bulk rows and then determine the coefficients from the boundary rows.

For the bulk rows $p+1,\ldots,L-q$, the eigenvalue equation is
\begin{equation}\label{BulkE2}
	\begin{aligned}
		E\,\psi(x) = T_{-p}\,\psi(x-p) + \cdots + T_{q}\,\psi(x+q),\\
		x=p+1,\dots,L-q.
	\end{aligned}
\end{equation} 
Substituting Eq.~(\ref{Psix}) into Eq.~(\ref{BulkE2}) gives
\begin{equation}
	\sum_{j=1}^{n_s} C_j\, (\beta_j)^x \left[ H(\beta_j) - E \right] u_j = 0.
\end{equation}
Each term vanishes by Eq.~(\ref{EigenFunction_n}); hence, the ansatz satisfies every bulk row for arbitrary coefficients $C_j$.

For the left boundary rows, we start from row $p$:
\begin{equation}\label{BoundaryL_p}
	E\,\psi(p) = T_{-p+1}\,\psi(1) + \cdots + T_{q}\,\psi(p+q).
\end{equation}
But extending the bulk recurrence~(\ref{BulkE2}) to row $p$ gives
\begin{equation}\label{BulkEp}
	\begin{aligned}
		E\,\psi(p) = & \, T_{-p}\,\psi(0)  \\
		& + T_{-p+1}\,\psi(1) + \cdots + T_{q}\,\psi(p+q),
	\end{aligned}
\end{equation}
Subtracting Eq.~(\ref{BoundaryL_p}) from Eq.~(\ref{BulkEp}) gives $T_{-p}\psi(0)=0$. Since $T_{-p}$ is invertible under the full-rank condition~(\ref{FRC}), this reduces to
\begin{equation}
	\psi(0) = 0.
\end{equation}
For row $p-1$, the extended bulk equation contains two terms absent from the OBC row, and their difference gives
\begin{equation}
	T_{-p}\,\psi(-1) + T_{-p+1}\,\psi(0) = 0.
\end{equation}
Using $\psi(0)=0$ and the invertibility of $T_{-p}$ reduces this equation to
\begin{equation}
	\psi(-1) = 0.
\end{equation}
Repeating this step successively for rows $p-2,\ldots,1$ yields the complete set of left boundary conditions
\begin{equation}
	\psi(x) = 0, \quad x=-p+1,\ldots,0.
\end{equation}

For the right boundary rows, the same recursion proceeds from row $L-q+1$ to row $L$. 
Since $T_q$ is invertible, it gives
\begin{equation}
	\psi(x) = 0, \quad x=L+1,\ldots,L+q.
\end{equation}
Here $x\in[-p+1,0]$ and $x\in[L+1,L+q]$ label fictitious lattice sites outside the physical chain.

Combining these results, the OBC truncation under the full-rank condition~(\ref{FRC}) is equivalent to
\begin{equation}
	\begin{cases} 
		\psi(x) = 0, & x \in [-p+1, 0] \quad (\text{Left BCs}) \\[1.5ex] \psi(x)= 0, & x \in [L+1,L+q] \quad (\text{Right BCs}) 
	\end{cases}
\end{equation}
The left and right boundaries yield $np$ and $nq$ constraints, respectively, giving a total of $n_s=n(p+q)$, equal to the number of undetermined coefficients in Eq.~(\ref{Psix}). Substituting Eq.~(\ref{Psix}) then gives the homogeneous system for $\mathbf{C}=[C_1,\dots,C_{n_s}]^T$:
\begin{equation}\label{MBC}
	M_B(E) ~\mathbf{C} = 0,
\end{equation} 
where the $n_s \times n_s$ boundary matrix $M_B(E)$ takes the form
\begin{equation}\label{BME}
	M_B(E)\!=\!
	\begin{pmatrix}
		(\beta_1)^{-p+1}\, u_1 & \cdots & (\beta_{n_s})^{-p+1}\, u_{n_s}\tikzmark{mbLtop} \\
		\vdots & \vdots & \vdots \\
		(\beta_1)^{0}\, u_1 & \cdots & (\beta_{n_s})^{0}\, u_{n_s}\tikzmark{mbLbot} \\[0.5ex] \hline\\[-2ex]
		(\beta_1)^{L+1}\, u_1 & \cdots & (\beta_{n_s})^{L+1}\, u_{n_s}\tikzmark{mbRtop} \\
		\vdots & \vdots & \vdots \\
		(\beta_1)^{L+q}\, u_1 & \cdots & (\beta_{n_s})^{L+q}\, u_{n_s}\tikzmark{mbRbot}
	\end{pmatrix}\hspace{45pt}
\end{equation}
\begin{tikzpicture}[overlay, remember picture]
	\coordinate (Ltop)  at ([xshift=11pt, yshift=6pt]pic cs:mbLtop);
	\coordinate (pLbot) at (pic cs:mbLbot);
	\coordinate (Lbot)  at ([yshift=-2pt]Ltop |- pLbot);
	\draw[decorate, decoration={brace, amplitude=4pt}, thick] (Ltop) -- (Lbot) node[midway, right=5pt] {\small Left BCs};
	
	\coordinate (Rtop)  at ([xshift=13pt, yshift=7pt]pic cs:mbRtop);
	\coordinate (pRbot) at (pic cs:mbRbot);
	\coordinate (Rbot)  at ([yshift=-2pt]Rtop |- pRbot);
	\draw[decorate, decoration={brace, amplitude=4pt}, thick] (Rtop) -- (Rbot) node[midway, right=5pt] {\small Right BCs};
\end{tikzpicture}
Each entry $(\beta_j)^x u_j$ is an $n$-component column vector, so $M_B(E)$ is an $n_s\times n_s$ matrix.
	Its upper $np$ rows impose the left boundary conditions through the factors $\beta^{-p+1},\dots,\beta^0$ and have no explicit $L$ dependence at fixed $E$.
	The lower $nq$ rows impose the right boundary conditions through $\beta^{L+1},\dots,\beta^{L+q}$.
	Their explicit $L$ dependence accounts for propagation across the chain.

Since $M_B(E)$ is square, Eq.~(\ref{MBC}) admits a nonzero solution $\mathbf{C}$ if and only if
\begin{equation}\label{EQC}
	\det [M_B(E)] = 0.
\end{equation}
This is an exact OBC eigenvalue condition when $M_B(E)$ is constructed from a complete, linearly independent basis of bulk solutions.
For an OBC eigenenergy $E$, we first solve the ChE~(\ref{ChE}) for $\{\beta_j(E)\}$ and Eq.~(\ref{EigenFunction_n}) for the internal vectors $\{u_j\}$.
We then find the null space of $M_B(E)$ in Eq.~(\ref{MBC}) to determine $\{C_j\}$.
Substituting these coefficients into Eq.~(\ref{Psix}) gives the OBC wavefunction.

\subsection{\texorpdfstring{Questions in the Thermodynamic Limit}{Questions in the Thermodynamic Limit}}

For a specified $H(\beta)$ and the OBC termination in Eq.~(\ref{HOBC}), we ask five questions about the thermodynamic limit:

\noindent(i) Spectrum coverage.\enspace
What is the set of OBC spectral accumulation points in the complex energy plane as $L\to\infty$?
We define this set by
\begin{equation}\label{sigmaOBC}
	\sigma_{\mathrm{OBC}}^{\mathrm{1D}} := \lim_{L\to\infty} \mathrm{spec}\big(H_{\mathrm{OBC}}(L)\big).
\end{equation}
This set may include isolated edge-state energies in addition to the bulk spectral support described by the GBZ.

\noindent(ii) Density of states.\enspace
What is the corresponding density of states (DOS) in the thermodynamic limit?
We define it by
\begin{equation}\label{DOS1}
	\rho_{\mathrm{OBC}}(E) := \lim_{L\to\infty} \frac{1}{nL} \sum_{i=1}^{nL} \delta^{(2)}(E - E_i),
\end{equation}
where $\{E_i\}\in \sigma_{\mathrm{OBC}}^{\mathrm{1D}}$ are counted with algebraic multiplicity.
Here $\delta^{(2)}(z)=\delta(\Re z)\delta(\Im z)$ is the Dirac distribution on the complex plane.
A finite number of edge states contributes vanishing normalized weight as $L\to\infty$.

\noindent(iii) Wavefunction profile.\enspace
For each chain, let $E$ denote an eigenenergy. 
For distinct roots, the wavefunction takes the form
\begin{equation}\label{PsiE_ansatz}
	\psi_E(x) = \sum_{j=1}^{n_s} C_j\, (\beta_j)^x\, u_j.
\end{equation}
How do the coefficients $\{C_j\}$ behave as $L\to\infty$?

\noindent(iv) Green's function.\enspace
Let $|i,\alpha\rangle$ denote orbital $\alpha$ in unit cell $i$.
We define the orbital-resolved Green's function by
\begin{equation}\label{GF_split}
	G_{\alpha\beta}(i,j;E)
	= \langle i,\alpha |
	[E\mathbbm{I}_{nL}-H_{\rm OBC}(L)]^{-1}
	|j,\beta\rangle.
\end{equation}
What is the analytical expression for this matrix element as $L\to\infty$?

\noindent(v) Finite-size criterion.\enspace
For a given $H(\beta)$, can we determine a critical size $L_c$ beyond which the finite-size spectrum approximates its thermodynamic limit?

\section{Exact Calculations for Finite Systems}\label{sec:Exact Calculations}

Before taking the thermodynamic limit $L \to \infty$, we derive finite-size quantization equations from the BME.
For a chain of length $L$, the OBC eigenenergies and their associated roots $\{\beta_j\}_{j=1}^{n_s}$ satisfy the BME $\det [M_B(E)] = 0$ [Eq.~(\ref{EQC})].
If $E$ and the $n_s$ roots are treated as independent variables, Eq.~(\ref{EQC}) must be supplemented by $n_s$ algebraic relations.
Vieta's formulas provide these relations by connecting the characteristic roots to the polynomial coefficients.
Under the full-rank condition, multiplying the ChE~(\ref{ChE}) by $\beta^{np}$ gives a degree-$n_s$ polynomial equation:
\begin{equation}
	\begin{aligned}
		P(\beta, E) &= \beta^{np}\det[E\mathbbm{I}_n - H(\beta)]= \sum_{k=0}^{n_s} a_k(E) \beta^k \\
		&= a_{n_s}(E) \prod_{j=1}^{n_s} [\beta - \beta_j(E)] = 0,
	\end{aligned}
\end{equation}
where the coefficients $a_k(E)$ depend on the hopping matrices $\{T_i\}$ and the energy $E$.
Vieta's formulas relate these coefficients to the roots of the characteristic polynomial:
\begin{widetext}
\begin{equation}\label{Vieta}
	\sum_{i} \beta_i = - \frac{a_{n_s-1}(E)}{a_{n_s}(E)}, \quad
	\sum_{i<j} \beta_i \beta_j = \frac{a_{n_s-2}(E)}{a_{n_s}(E)}, \quad
	\dots, \quad
	\prod_{i} \beta_i = (-1)^{n_s} \frac{a_0(E)}{a_{n_s}(E)}.
\end{equation}
\end{widetext}
Together with the eigenvalue condition~(\ref{EQC}), these give $n_s+1$ complex equations for the $n_s+1$ unknowns $\{\beta_j\}$ and $E$.
Their solutions determine the OBC eigenvalues; the corresponding null vectors of $M_B(E)$ determine the eigenstates.

For small $n_s$, we can eliminate all but one characteristic root to obtain a single-variable algebraic equation.
In each elimination below, $\beta$ denotes a selected root, temporarily labeled $\beta_1$ independently of the modulus ordering.
We illustrate this procedure with two examples: a single-band model satisfying the full-rank condition~(\ref{FRC}) (Sec.~\ref{subsec:single_band_BME}), and a non-Hermitian SSH model that violates it (Sec.~\ref{subsec:SSH_BME}).

\subsection{Example: Single-Band Model with Long-Range Hopping}\label{subsec:single_band_BME}

As a first example, consider a single-band tight-binding model ($n=1$) with maximum hopping ranges $p=2$ and $q=1$.
This model satisfies the full-rank condition and its non-Bloch Hamiltonian reads
\begin{equation}
	H(\beta) = t_1 \beta + t_0 + \frac{t_{-1}}{\beta} + \frac{t_{-2}}{\beta^2}.
\end{equation}
At a given energy $E$, the ChE $H(\beta)=E$ becomes the cubic equation
\begin{equation}\label{single_band_ChE}
	t_1 \beta^3 + (t_0 - E) \beta^2 + t_{-1} \beta + t_{-2} = 0,
\end{equation}
with three roots $\beta_1, \beta_2, \beta_3$, counted with multiplicity. 

Since $n=1$, the internal eigenvectors can be chosen as unity.
For three distinct roots, the $n_s=3$ fictitious-site conditions give the boundary matrix [cf.\ Eq.~(\ref{MBC})]
\begin{equation}
	M_B(E) = \begin{pmatrix}
		(\beta_1)^{-1} & (\beta_2)^{-1} & (\beta_3)^{-1} \\[1ex]
		1 & 1 & 1 \\[1ex]
		(\beta_1)^{L+1} & (\beta_2)^{L+1} & (\beta_3)^{L+1}
	\end{pmatrix}.
\end{equation}
Expanding the determinant along the first row gives
\begin{equation}
	\begin{split}
	\det [M_B(E)]  = & + (\beta_1)^{-1}\bigl[(\beta_3)^{L+1} - (\beta_2)^{L+1}\bigr] \\
	&- (\beta_2)^{-1}\bigl[(\beta_3)^{L+1} - (\beta_1)^{L+1}\bigr] \\
	&+ (\beta_3)^{-1}\bigl[(\beta_2)^{L+1} - (\beta_1)^{L+1}\bigr].
	\end{split}
\end{equation}
Collecting the terms proportional to $(\beta_i)^{L+1}$ and extracting the common factor $(\beta_1\beta_2\beta_3)^{-1}$, we obtain
\begin{equation}
	\begin{split}
	\det[ M_B(&E)] = \frac{1}{\beta_1\beta_2\beta_3}\bigl[
	(\beta_1)^{L+2}(\beta_3 - \beta_2) \\
	&+(\beta_2)^{L+2}(\beta_1 - \beta_3)+ (\beta_3)^{L+2}(\beta_2 - \beta_1)\bigr].
	\end{split}
\end{equation}
Since $\beta_1\beta_2\beta_3 = -t_{-2}/t_1 \neq 0$ by assumption, the condition $\det[ M_B(E)] = 0$ reduces to
\begin{equation}
	(\beta_1)^{L+2}(\beta_3 - \beta_2) + (\beta_2)^{L+2}(\beta_1 - \beta_3) + (\beta_3)^{L+2}(\beta_2 - \beta_1) = 0.
\end{equation}
Dividing by $\beta_3 - \beta_2$ (assuming $\beta_2 \neq \beta_3$) and rearranging, we obtain
\begin{equation}\label{single_band_det_div}
	\begin{aligned}
	(\beta_1)^{L+2} ={}& \beta_1 \frac{(\beta_2)^{L+2}-(\beta_3)^{L+2}}{\beta_2 - \beta_3}\\
	&-\beta_2\beta_3\frac{(\beta_2)^{L+1}-(\beta_3)^{L+1}}{\beta_2-\beta_3}.
	\end{aligned}
\end{equation}
Vieta's formulas for the cubic~(\ref{single_band_ChE}) read
\begin{equation}\label{single_band_Vieta}
	\begin{aligned}
		\beta_1 + \beta_2 + \beta_3 &= \frac{E - t_0}{t_1}, \\
		\beta_1\beta_2 + \beta_2\beta_3 + \beta_1\beta_3 &= \frac{t_{-1}}{t_1}, \\
		\beta_1\beta_2\beta_3 &= -\frac{t_{-2}}{t_1},
	\end{aligned}
\end{equation}
and Eqs.~(\ref{single_band_det_div}) and~(\ref{single_band_Vieta}) provide the starting point for the elimination.
Eliminating the auxiliary roots $\beta_2$ and $\beta_3$ from Eq.~(\ref{single_band_det_div}) leaves a single equation for the selected root $\beta_1 \equiv \beta$.

In Vieta's formulas~(\ref{single_band_Vieta}), the third and second relations give the product $v(\beta)$ and sum $u(\beta)$ of the remaining roots:
\begin{equation}\label{single_band_uv}
	\begin{aligned}
		v(\beta) \equiv \beta_2 \beta_3 &= -\frac{t_{-2}}{t_1 \beta}, \\
		u(\beta) \equiv \beta_2 + \beta_3 &= \frac{1}{\beta} \left( \frac{t_{-1}}{t_1} - v(\beta) \right) \\
		&= \frac{t_{-1}\beta + t_{-2}}{t_1 \beta^2}.
	\end{aligned}
\end{equation}
The auxiliary roots $\beta_2$ and $\beta_3$ therefore satisfy 
\begin{equation}
	z^2-u(\beta)z+v(\beta)=0.
\end{equation}
Solving this quadratic gives
\begin{equation}
	\beta_{2,3} = \frac{u(\beta) \pm \sqrt{\Delta(\beta)}}{2},
\end{equation}
where $\Delta(\beta)\equiv u^2(\beta)-4v(\beta)$ is the discriminant.
To rewrite Eq.~(\ref{single_band_det_div}), we define
\begin{equation}\begin{aligned}
	D_k(\beta)& = \frac{\beta_2^{k+1} - \beta_3^{k+1}}{\beta_2 - \beta_3}\\
	&=\frac{1}{\sqrt{\Delta}} \Biggl[ \left( \frac{u + \sqrt{\Delta}}{2} \right)^{\!k+1} - \left( \frac{u - \sqrt{\Delta}}{2} \right)^{\!k+1} \Biggr].
\end{aligned}\end{equation}
Because $u(\beta)$ and $v(\beta)$ depend only on $\beta$ [Eq.~(\ref{single_band_uv})], so does $D_k(\beta)$. 

\begin{figure}[b]
	\begin{center}
		\includegraphics[width=0.9\linewidth]{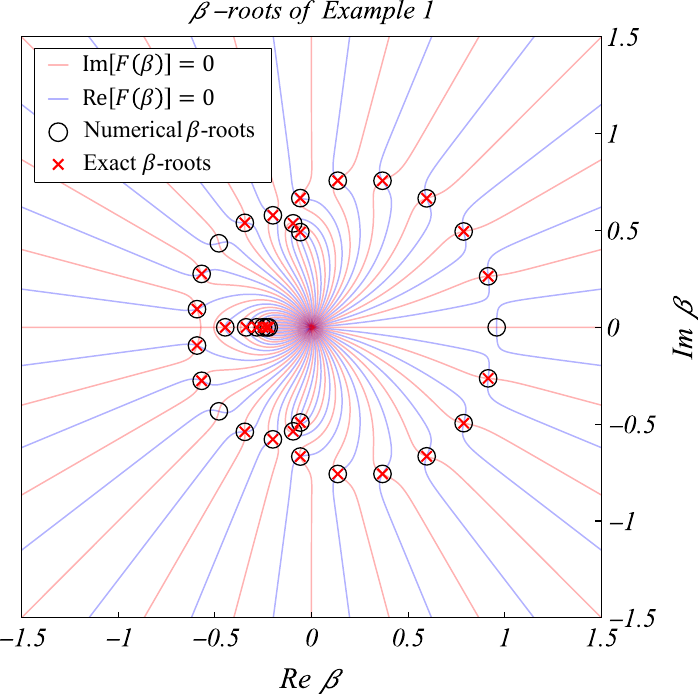}
	\end{center}
	\protect\caption{Algebraic quantization for the single-band model. Blue and red curves show the contours $\mathrm{Re}[F(\beta)]=0$ and $\mathrm{Im}[F(\beta)]=0$, respectively. Their intersections are marked by circles. Crosses denote the roots obtained by substituting the eigenvalues from exact diagonalization into the ChE~(\ref{single_band_ChE}). The crosses coincide with the circles. Unmatched circles arise from $\beta_i=\beta_j$ where $i\neq j$. The parameters are $t_1=1.0$, $t_0=0$, $t_{-1}=0.5$, $t_{-2}=0.2$, and $L=10$. All computations use 16-digit numerical precision.}
	\label{Fig_single_band_crossings}
\end{figure}

We can now write Eq.~(\ref{single_band_det_div}) in terms of $D_k(\beta)$ as
\begin{equation}
	\beta^{L+2} = \beta\, D_{L+1}(\beta) - v(\beta)\, D_L(\beta).
\end{equation}
Rearranging this expression gives the algebraic quantization equation
\begin{equation}\label{single_band_quantization}
	F(\beta) \equiv \beta^{L+2} + v(\beta)\, D_L(\beta) - \beta\, D_{L+1}(\beta) = 0.
\end{equation}
Combining the BME with Vieta's formulas has thus reduced the quantization condition to a single algebraic equation in $\beta$.

To verify this result, we choose $t_1=1$, $t_0=0$, $t_{-1}=0.5$, $t_{-2}=0.2$, and $L=10$.
Fig.~\ref{Fig_single_band_crossings} shows the contours $\mathrm{Re}[F(\beta)]=0$ (blue) and $\mathrm{Im}[F(\beta)]=0$ (red) in the complex-$\beta$ plane.
Their intersections, marked by circles, give the numerical roots of $F(\beta)=0$.
For comparison, we diagonalize the OBC Hamiltonian and substitute its eigenvalues into the ChE~(\ref{single_band_ChE}).
The resulting roots, marked by crosses, agree with the physical solutions of the quantization equation.

For the plotted parameters, the unmatched circles correspond to repeated characteristic roots, with $\beta_2=\beta_3$ after ordering by modulus.
This violates the distinct-root assumption used in our derivation: two columns of the boundary matrix become identical, making its determinant vanish automatically.
Numerical diagonalization confirms that the corresponding energies are not OBC eigenvalues, so these points are spurious solutions.

\subsection{Example: Non-Hermitian SSH Model with Rank Deficiency}\label{subsec:SSH_BME}

We next apply the BME to a model that violates the full-rank condition~(\ref{FRC}).
Consider the non-Hermitian SSH model with nonreciprocal intracell hopping.
This two-band model ($n=2$) has nearest-neighbor intercell hopping ($p=q=1$), and its non-Bloch Hamiltonian reads
\begin{equation}
	H(\beta) = \begin{pmatrix} 0 & t_1+\delta+t_2/\beta \\ t_1-\delta+t_2\beta & 0 \end{pmatrix}.
\end{equation}
The corresponding hopping matrices are
\begin{equation}\label{SSH_Ti}
	T_0 = \begin{pmatrix} 0 & t_1+\delta \\ t_1-\delta & 0 \end{pmatrix}, 
	T_1 = \begin{pmatrix} 0 & 0 \\ t_2 & 0 \end{pmatrix}, 
	T_{-1} = \begin{pmatrix} 0 & t_2 \\ 0 & 0 \end{pmatrix}.
\end{equation}
Both $T_1$ and $T_{-1}$ are rank-deficient ($\det T_{\pm1}=0$), violating the full-rank condition~(\ref{FRC}).
For $t_2(t_1+\delta)(t_1-\delta)\neq0$, the determinant $\det[E\mathbbm{I}_2-H(\beta)]$ has a first-order pole at $\beta=0$.
Therefore, in the following discussion, we assume $t_1\neq \pm \delta$. 
Multiplication by $\beta$ therefore gives a quadratic equation, rather than the degree-$4$ polynomial expected under full rank:
\begin{equation}
	t_2(t_1+\delta)\beta^2 + [(t_1)^2-\delta^2+(t_2)^2-E^2]\beta + t_2(t_1-\delta) = 0.
\end{equation}
The quadratic has two nonzero roots, $\beta_1$ and $\beta_2$, at each energy $E$.
Vieta's formulas give their sum and product:
\begin{equation}
	\begin{aligned}
		\beta_1+\beta_2 &= \frac{E^2-t_1^2+\delta^2-t_2^2}{t_2(t_1+\delta)}, \\
		\beta_1\beta_2 &= \frac{t_1-\delta}{t_1+\delta} \equiv r.
	\end{aligned}
\end{equation}
We construct the BME from these two roots. 

\begin{figure}[b]
	\begin{center}
		\includegraphics[width=0.9\linewidth]{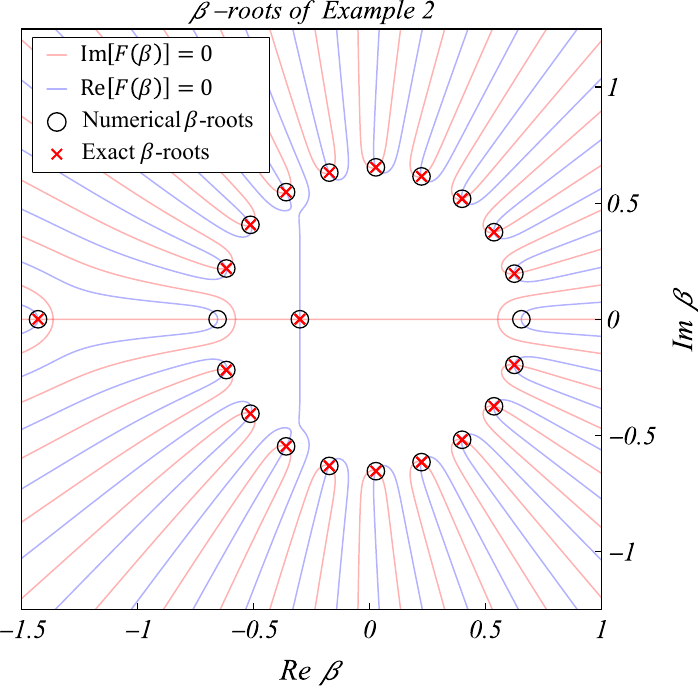}
	\end{center}
	\protect\caption{Algebraic quantization for the non-Hermitian SSH model. Blue and red curves show the contours $\mathrm{Re}[F(\beta)]=0$ and $\mathrm{Im}[F(\beta)]=0$, respectively. Their intersections are marked by circles. Crosses denote the roots obtained by substituting the eigenvalues from exact diagonalization into the ChE. For these parameters, the physical roots describe bulk states on the GBZ circle $|\beta|=\sqrt{r}$ and topological edge states away from the circle. The quantization equation captures both types of states. The spurious roots $\beta=\pm\sqrt{r}$ occur where the two boundary-matrix columns coincide. The parameters are $t_1=0.5$, $\delta=0.2$, $t_2=1$, and $L=10$. All computations use 16-digit numerical precision.
	}
	\label{Fig_SSH_BME_results}
\end{figure}

For $\beta_1\neq\beta_2$, we write the bulk wavefunction $\psi(x)=[\psi_A(x),\psi_B(x)]^T$ as a superposition of two generalized Bloch modes:
\begin{equation}\label{SSH_ansatz}
	\psi(x) = \sum_{j=1}^2 C_j(\beta_j)^x u_j.
\end{equation} 
Here $u_j=(u_{j,A},u_{j,B})^T$ is an internal right eigenvector satisfying $H(\beta_j)u_j=Eu_j$.
For $E\neq0$, it can be chosen as
\begin{equation}
	u_j = \begin{pmatrix} E \\ t_1-\delta+t_2\beta_j \end{pmatrix}.
\end{equation}
The hopping matrices in Eq.~(\ref{SSH_Ti}) give the boundary conditions 
\begin{equation}
	\begin{cases}
		\psi_B(0) = 0 & (\text{Left BC}), \\[1ex]
		\psi_A(L+1) = 0 & (\text{Right BC}).
	\end{cases}
\end{equation}
Substituting the wavefunction ansatz into these fictitious-site conditions gives
\begin{equation}
	\begin{aligned}
		\psi_B(0) &= C_1(t_1-\delta+t_2\beta_1) + C_2(t_1-\delta+t_2\beta_2) = 0, \\
		\psi_A(L+1) &= C_1\, (\beta_1)^{L+1}E + C_2\, (\beta_2)^{L+1}E = 0.
	\end{aligned}
\end{equation}
We write these equations as $M_B(E)\,(C_1, C_2)^T = 0$, with 
\begin{equation}
	M_B(E) = \begin{pmatrix}
		t_1-\delta+t_2\beta_1 & t_1-\delta+t_2\beta_2 \\[1ex]
		(\beta_1)^{L+1}E &  (\beta_2)^{L+1}E
	\end{pmatrix}.
\end{equation}
Assuming $E$ and $t_1-\delta+t_2\beta_1$ are nonzero, we impose $\det[M_B(E)]=0$ and cancel these factors to obtain
\begin{equation}\label{SSH_Fbeta}
	\frac{t_1-\delta+t_2\beta_2}{t_1-\delta+t_2\beta_1} = \left(\frac{\beta_2}{\beta_1}\right)^{L+1}.
\end{equation}
Setting $\beta\equiv\beta_1$ and substituting $\beta_2=r/\beta$, we obtain the algebraic quantization equation
\begin{equation}\label{SSH_quantization}
	F(\beta) \equiv (t_1-\delta+t_2\,r/\beta)\beta^{2L+2}-(t_1-\delta+t_2\,\beta)r^{L+1}=0.
\end{equation}
The BME therefore yields a single-variable quantization equation even for this rank-deficient model.

To verify Eq.~(\ref{SSH_quantization}), we choose $t_1=0.5$, $\delta=0.2$, $t_2=1$, and $L=10$.
Fig.~\ref{Fig_SSH_BME_results} shows the contours $\mathrm{Re}[F(\beta)]=0$ (blue) and $\mathrm{Im}[F(\beta)]=0$ (red); circles mark their intersections.
For comparison, we diagonalize $H_{\mathrm{OBC}}$ and substitute its eigenvalues into the ChE to obtain the roots marked by crosses.
For these parameters, the physical roots describe bulk states on the GBZ circle $|\beta|=\sqrt{r}$ and topological edge states away from the circle.
The physical solutions of $F(\beta)=0$ agree with both the bulk and edge-state roots obtained from exact diagonalization. 

For the parameters used above, $F(\beta)=0$ also has two spurious roots at $\beta=\pm\sqrt{r}$.
At these points, $\beta_2=r/\beta_1=\beta_1$, violating the distinct-root assumption.
As in the preceding single-band example, the two boundary-matrix columns then coincide, making the determinant vanish automatically.

\section{Spectral Approach I: Single-Band Case}\label{sec:Spectral Approach}
The BME determines the OBC eigenvalue problem at finite system size $L$.
Combining $\det[M_B(E)]=0$ with Vieta's formulas gives explicit quantization equations for simple models.
However, eliminating the roots becomes difficult for more general models; therefore, it is natural to ask how can the spectrum be determined directly in the thermodynamic limit $L\to\infty$.
The spectral approach provides a way to solve this problem by analyzing the asymptotic behavior of $\det[M_B(E)]$ in the thermodynamic limit $L\to\infty$.

\subsection{GBZ Condition and Spectral Condensation}\label{subsec:GBZ Condition and Spectral Condensation}

We start with a 1D single-band model, for which the hopping matrices $T_i$ and internal eigenvectors reduce to scalars.
The boundary matrix in Eq.~(\ref{BME}) takes the form
\begin{equation}
	M_B(E)=
	\begin{pmatrix}
		(\beta_1)^{-p+1} & \cdots & (\beta_{p+q})^{-p+1} \\
		\vdots & \vdots & \vdots \\
		(\beta_1)^{0} & \cdots & (\beta_{p+q})^{0} \\[0.5 ex] \hline\\[-2ex]
		(\beta_1)^{L+1} & \cdots & (\beta_{p+q})^{L+1} \\
		\vdots & \vdots & \vdots \\
		(\beta_1)^{L+q} & \cdots & (\beta_{p+q})^{L+q}
	\end{pmatrix}.
\end{equation}
Expanding the determinant by the lower block selects $q$ columns, each contributing one factor $(\beta_j)^L$~{ \cite{Murakami2019PRL}}.
All remaining powers are independent of $L$ and enter a coefficient $A_{i_1\dots i_q}$.
Thus, the determinant can be written as
\begin{equation}
	\det[ M_B(E)] = \sum_{\{i_1, \dots, i_q\}} A_{i_1 \dots i_q} (\beta_{i_1} \beta_{i_2} \cdots \beta_{i_q})^L.
\end{equation}
Here the sum runs over $1\le i_1<\cdots<i_q\le p+q$.
With the root ordering in Eq.~(\ref{OrderBeta}), the largest and next-largest exponential factors are
\begin{equation}
	\begin{aligned}
		\Omega_{\max} &= (\beta_{p+1} \beta_{p+2} \cdots \beta_{p+q})^L\\
		\Omega_{\text{sub}} &= (\beta_{p} \beta_{p+2} \cdots \beta_{p+q})^L.
	\end{aligned}
\end{equation}
Dividing all terms by $\Omega_{\max}$ and setting the result to zero, the eigenvalue condition becomes
\begin{equation}\label{EQC_expansion}
	\begin{aligned}
		\frac{\det M_B(E)}{\Omega_{\max}}& = A_{\max} + A_{\text{sub}} \left( \frac{\beta_p}{\beta_{p+1}} \right)^L \\
		&+ \sum_{\text{others}} A_{i_1 \dots i_q} \left( \frac{\beta_{i_1} \dots \beta_{i_q}}{\beta_{p+1} \dots \beta_{p+q}} \right)^L = 0.
	\end{aligned}
\end{equation}
We now examine Eq.~(\ref{EQC_expansion}) in the thermodynamic limit $L\to\infty$.

If $|\beta_p|<|\beta_{p+1}|$, all terms after $A_{\max}$ in Eq.~(\ref{EQC_expansion}) vanish exponentially as $L\to\infty$, giving
\begin{equation}
	\lim_{L\to\infty}\frac{\det M_B(E)}{\Omega_{\max}}
		= A_{\max}(E).
\end{equation}
For distinct roots in the scalar model, $A_{\max}(E)$ is nonzero, so the strict-gap region contains no continuous bulk spectrum.

Next, if $|\beta_{p-1}|<|\beta_p|=|\beta_{p+1}|<|\beta_{p+2}|$, only the two leading terms in Eq.~(\ref{EQC_expansion}) survive as $L\to\infty$, giving
\begin{equation}\label{standing_wave_balance}
	\left(\frac{\beta_{p+1}(E)}{\beta_p(E)}\right)^L
		\simeq -\frac{A_{\mathrm{sub}}(E)}{A_{\max}(E)}.
\end{equation}
With $\Theta(E)\equiv\arg[-A_{\mathrm{sub}}(E)/A_{\max}(E)]$, taking the logarithm of Eq.~(\ref{standing_wave_balance}) introduces an integer $m\in\mathbb Z$:
\begin{equation}\label{standing_wave_log}
	\begin{aligned}
		\log\!\left[\frac{\beta_{p+1}(E)}{\beta_p(E)}\right]
		&\simeq \frac{1}{L}\log\!\left|\frac{A_{\mathrm{sub}}(E)}{A_{\max}(E)}\right|\\
		&\quad+\frac{i}{L}\left[2\pi m+\Theta(E)\right].
	\end{aligned}
\end{equation}
The real part gives $\log|\beta_{p+1}/\beta_p|=O(L^{-1})$, a finite-size deviation.
With locally continuous $k_j(E)=\arg\beta_j(E)$ ($j=p,p+1$) and $\Delta k(E)\equiv k_{p+1}(E)-k_p(E)$, the imaginary part at the eigenvalue $E_m$ gives
\begin{equation}\label{standing_wave_phase}
	\Delta k(E_m)\simeq\frac{2\pi m+\Theta(E_m)}{L}.
\end{equation}
On a regular spectral arc, there is $|E_{m+1}-E_m|=O(L^{-1})$.
Smoothness of $\Theta$ then gives
\begin{equation}\label{phase_variation}
	\Theta(E_{m+1})-\Theta(E_m)= O\!\left(|E_{m+1}-E_m|\right)=O(L^{-1}).
\end{equation}
Subtracting Eq.~(\ref{standing_wave_phase}) for adjacent $m$ yields
\begin{equation}
	\begin{aligned}
		\delta(\Delta k)&\equiv\left|\Delta k(E_{m+1})-\Delta k(E_m)\right|\\
		&\simeq\frac{1}{L}\left|2\pi+\Theta(E_{m+1})-\Theta(E_m)\right|\\
		&=\frac{2\pi}{L}+O(L^{-2})\xrightarrow{L\to\infty}0.
	\end{aligned}
\end{equation}
Thus, the standing waves become increasingly dense, and their energies condense onto a continuous spectral arc as $L\to\infty$.

\begin{figure}[t]
	\centering
	\includegraphics[width=1\linewidth]{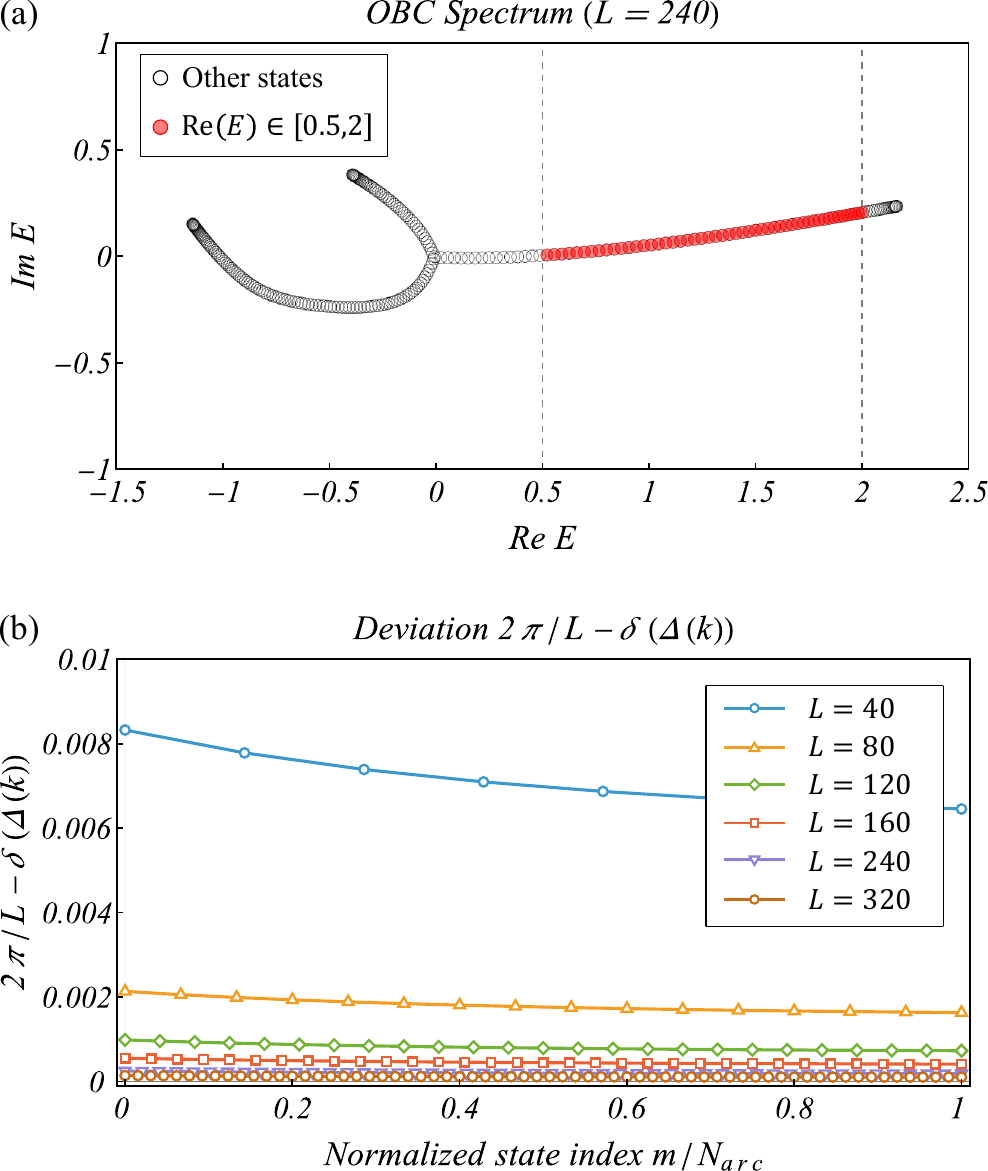}
	\caption{Momentum spacing for a 1D single-band non-Hermitian model with $p=2$ and $q=3$. (a) OBC spectrum at $L=240$; red points mark the selected arc $\mathrm{Re}(E)\in[0.5,2]$. (b) Deviation $2\pi/L-\delta(\Delta k)$ for $L=40,80,120,160,240$, and $320$, plotted against the normalized state index $m/N_{\mathrm{arc}}$, where $N_{\mathrm{arc}}$ counts the selected states. The convergence is consistent with $2\pi/L-\delta(\Delta k)=O(1/L^2)$. Parameters: $(t_{-2},t_{-1},t_0,t_1,t_2,t_3)=(0.4+0.1i,0.7-0.15i,0.1+0.05i,0.5+0.2i,0.3-0.08i,0.15+0.12i)$. All computations use 50-digit numerical precision.}
	\label{fig:delta_k_spacing}
\end{figure}

As a concrete example, we consider the single-band model $H(\beta)=\sum_{j=-2}^{3}t_j\beta^j$ with hopping ranges $p=2$ and $q=3$.
The parameters are $t_{-2}=0.4+0.1i$, $t_{-1}=0.7-0.15i$, $t_0=0.1+0.05i$, $t_1=0.5+0.2i$, $t_2=0.3-0.08i$, and $t_3=0.15+0.12i$.
Fig.~\ref{fig:delta_k_spacing}(a) shows the OBC spectrum obtained by numerical diagonalization at $L=240$.
We select the arc with $\mathrm{Re}(E)\in[0.5,2.0]$ and plot $2\pi/L-\delta(\Delta k)$ in Fig.~\ref{fig:delta_k_spacing}(b) for $L=40,80,120,160,240$, and $320$.
The decreasing deviation is consistent with the analytical convergence.

In summary, the GBZ for a single-band model is defined as
\begin{equation}\label{GBZ_def}
	\beta_{\text{GBZ}}^{\text{1D}}
		= \bigcup_{\substack{E \in \mathbb{C} \\ |\beta_{p+1}(E)| = |\beta_p(E)|}}
		\!\bigl\{\beta_p(E),\beta_{p+1}(E)\bigr\}.
\end{equation}
Since neighboring states differ by $\delta m=1$, the number of states per unit phase difference is $\delta m/\delta(\Delta k)=[\delta(\Delta k)]^{-1}=L/(2\pi)+O(1)$.
We will use this relation below to obtain the DOS along the OBC spectrum.

\begin{table*}[t]
	\centering
	\caption{Equivalent formulations of the GBZ condition for one-dimensional single-band systems.}
	\label{tab:gbz_criteria}
	\begin{tabular}{llll}
		\toprule
		Method & Core criterion & $E \in \sigma_{\text{GBZ}}^{\text{1D}}$ & $E \notin \sigma_{\text{GBZ}}^{\text{1D}}$ \\
		\midrule
		Root criterion & Ordering of $\beta$ & $|\beta_p| = |\beta_{p+1}|$ & $|\beta_p| \neq |\beta_{p+1}|$ \\
		\addlinespace
		Amoeba & Logarithmic central hole & No central hole & Presence of central hole \\
		\addlinespace
		Winding number & Topology of reference circle & No zero-winding radius & A zero-winding radius exists \\
		\addlinespace
		Spectral potential & Harmonicity of $\Phi(E)$ & Not locally harmonic & Locally harmonic \\
		\bottomrule
	\end{tabular}
\end{table*}

\subsection{Predictions and Limitations of the GBZ}\label{subsec:Predictions and Limitations of the GBZ}

We now use the GBZ to answer the five questions posed in Sec.~\ref{sec:BME}.

\noindent(i) Spectrum coverage.\enspace
The continuous bulk OBC spectrum is the image of the GBZ under the Hamiltonian,
\begin{equation}
	\sigma_{\text{GBZ}}^{\text{1D}}=H(\beta_{\text{GBZ}}^{\text{1D}}).
\end{equation} 
Thus, evaluating $H(\beta)$ on the GBZ gives the thermodynamic bulk spectrum. In the Hermitian single-band limit, the GBZ becomes the unit circle and the bulk OBC and PBC spectra both reduce to $H(e^{ik})$.

\noindent(ii) Density of states.\enspace
Let $s$ denote arc length along a regular spectral arc ($ds=|dE|$), and $\delta s$ the spacing between adjacent states.
Using the state-counting relation above, the DOS per unit cell and unit arc length is
\begin{equation}\label{DOS}
	\begin{aligned}
		\rho_{\mathrm{OBC}}^{\mathrm{1D}}(E(s))
		&\equiv\lim_{L\to\infty}\frac{1}{L}\frac{\delta m}{\delta s}\\
		&=\lim_{L\to\infty}\frac{1}{L}
		\frac{\delta m}{\delta(\Delta k)}
		\frac{\delta(\Delta k)}{\delta s}\\
		&=\frac{1}{2\pi}\left|\frac{d\Delta k}{ds}\right|\\
		&=\frac{1}{2\pi}\left|\frac{dk_{p+1}}{ds}-\frac{dk_p}{ds}\right|.
	\end{aligned}
\end{equation}
If the two phases vary in opposite directions along the arc, Eq.~(\ref{DOS}) reduces to $\rho_{\mathrm{OBC}}^{\mathrm{1D}}=(2\pi)^{-1}\sum_{j=p}^{p+1}|dE/dk_j|^{-1}$.
When additional roots share the GBZ modulus, the two-term approximation must be extended.
In the Hermitian limit, counting all propagating momenta gives the standard van Hove formula
\begin{equation}
	\rho_{\text{Hermitian}}(E)
		=\frac{1}{2\pi}\sum_i\frac{1}{|v(k_i)|}.
\end{equation}
Here $k_i$ runs over momenta satisfying $H(e^{ik_i})=E$, and $v(k)\equiv dH(e^{ik})/dk$ is the group velocity.

\noindent(iii) Wavefunction profile.\enspace
The GBZ fixes the leading exponential envelope of an OBC eigenstate,
\begin{equation}
	|\psi(x)|\sim e^{\mu_E x},\qquad
		\mu_E=\ln|\beta_p(E)|=\ln|\beta_{p+1}(E)|,
\end{equation} 
where the sign of $\mu_E$ determines the localization direction. However, the GBZ condition does not determine the detailed wavefunction, which will be investigated in future work in this series.

\noindent(iv) Green's function.\enspace
The GBZ determines the leading spatial decay of the two-point Green's function $G(x,x';E)$. For large separations $|x-x'|\gg 1$, it gives~\cite{Wentan2021PRB}
\begin{equation}
	G(x,x';E)\sim
		\begin{cases}
			[\beta_p(E)]^{x-x'}, & x>x',\\[1ex]
			[\beta_{p+1}(E)]^{x-x'}, & x<x'.
	\end{cases}
\end{equation} 
The two roots describe propagation in opposite directions and thus capture the non-reciprocal response. However, the spectral approach determines only these decay exponents. The exact superposition coefficients and prefactors will be addressed in future work in this series.

\noindent(v) Finite-size criterion.\enspace
The spectral approach gives the thermodynamic spectrum but no finite-size corrections or criterion for $L_c$. This limitation is especially important for the critical NHSE, where convergence can be anomalously slow~\cite{Zhesen2020PRL,Linhu2020NC}. Future work in this series will use coefficient flow to describe this crossover and its finite-size scaling.

\section{Spectral Approach II: Alternative Formulations of the GBZ Condition}\label{subsec:Alternative Formulations}

The preceding analysis connects the GBZ condition $|\beta_p| = |\beta_{p+1}|$ to the OBC spectrum, DOS, wavefunction envelope and Green's function in the thermodynamic limit.
A practical question is then: given a trial complex energy $E \in \mathbb{C}$, how can one determine whether it belongs to the continuous OBC spectrum $\sigma_{\text{GBZ}}^{\text{1D}}$?
Several formulations address this problem through characteristic roots, amoeba, winding numbers, or spectral potentials.
These formulations provide algebraic, geometric, topological, and analytical descriptions of the same GBZ condition in 1D single-band systems.

Throughout this section, we consider 1D single-band models ($n=1$).
The extension to multi-band systems ($n>1$) introduces additional constraints, as discussed in Sec.~\ref{sec:Limitations and Challenges of the Spectral Approach}.

\subsubsection{Root Criterion}

The algebraic root criterion provides the most direct test of spectra~\cite{Murakami2019PRL}.
For a trial energy $E$, we solve the ChE $f(E,\beta)=0$ and order its $p+q$ roots by modulus, $|\beta_1|\le\dots\le|\beta_{p+q}|$.
The criterion reads
\begin{equation}
	E \in \sigma_{\text{GBZ}}^{\text{1D}} \iff |\beta_p(E)| = |\beta_{p+1}(E)|.
\end{equation}
A strict gap $|\beta_p(E)|<|\beta_{p+1}(E)|$ therefore excludes $E$ from the continuous OBC spectrum.
This is precisely the GBZ condition derived in Sec.~\ref{sec:Spectral Approach}.

\subsubsection{Amoeba Criterion}

The amoeba formulation expresses root ordering through the logarithmic moduli of the characteristic roots~\cite{Hongyi2024PRX}.
We define $\mu_j\equiv\ln|\beta_j|$, which gives the ordered sequence
\begin{equation}
	\mu_1 \le \mu_2 \le \dots \le \mu_{p+q}.
\end{equation}
The set of these logarithmic coordinates is the 1D amoeba of $f(E,\beta)$ at fixed $E$.
The spectrum is determined by the central logarithmic gap:
\begin{equation}
	\mu_{p+1}(E) - \mu_p(E) = \ln\frac{|\beta_{p+1}|}{|\beta_p|} \begin{cases} > 0 & \Rightarrow E \notin \sigma_{\text{GBZ}}^{\text{1D}}, \\[1ex] = 0 & \Rightarrow E \in \sigma_{\text{GBZ}}^{\text{1D}}. \end{cases}
\end{equation}
Geometrically, the open interval $(\mu_p,\mu_{p+1})$ is the central hole in the complement of the amoeba.
Its closure reproduces the root criterion for the continuous OBC spectrum.
The geometric construction also extends to higher dimensions, where the topology of the amoeba provides information about complex spectra.

\subsubsection{Winding Number Criterion}

The winding number formulation gives a topological version of the same test~\cite{Kai2020PRL,Okuma2020PRL}:
Can a circle in the complex-$\beta$ plane separate the inner $p$ roots from the outer $q$ roots?
For a trial energy $E$ and a radius $R>0$, provided $f(E,\beta)\neq0$ on the counterclockwise contour $|\beta|=R$, the spectral winding number is defined as
\begin{equation}
	\begin{aligned}
		w(E, R) &\equiv \oint_{|\beta|=R} \frac{d\beta}{2\pi i} \frac{\partial}{\partial \beta} \ln \det(H(\beta) - E)\\
		&= N_{\text{zeros}}(<R) - N_{\text{poles}}(<R),
	\end{aligned}
\end{equation}
where $N_{\text{zeros}}(<R)$ counts the roots satisfying $|\beta|<R$, including multiplicity, and $N_{\text{poles}}(<R)=p$ for 1D single-band models.
The spectral membership criterion reads
\begin{equation}
	\begin{aligned}
	E \in \sigma_{\text{GBZ}}^{\text{1D}} &\iff \nexists\,\text{admissible }R>0\text{ with }w(E,R)=0 \\
	&\iff \nexists\,R>0\text{ with }|\beta_p|<R<|\beta_{p+1}|.
	\end{aligned}
\end{equation}
When $|\beta_p|=|\beta_{p+1}|$, no admissible circle encloses exactly $p$ roots.
Therefore, no zero-winding radius exists, in agreement with the root criterion.

\subsubsection{Spectral Potential Criterion}

The spectral potential formulation relates complex energy spectra to charge distributions in the two-dimensional energy plane.
For the single-band models considered here, the potential is determined by the $q$ largest-modulus roots~\cite{Yuncheng2024PRB}:
\begin{equation}
	\Phi(E) = \sum_{j=p+1}^{p+q} \ln |\beta_j(E)| + \ln|T_q|.
\end{equation}
Here $T_q$ is the hopping coefficient in $H(\beta)$. 
Outside the continuous OBC spectrum, the potential satisfies Laplace's equation in a neighborhood of $E$:
\begin{equation}
	\nabla^2 \Phi(E) = 4\, \partial_E \partial_{\bar{E}}\, \Phi(E) = 0,
\end{equation}
where $\nabla^2$ acts on the real and imaginary parts of $E$, and $\partial_E$, $\partial_{\bar E}$ are the corresponding Wirtinger derivatives.
A strict root gap persists under small energy perturbations, so the set of $q$ dominant roots remains separated from the remaining roots.
Its product is locally analytic and nonzero, making the sum of logarithmic moduli harmonic.
On the spectrum, the Laplacian must be interpreted as a distribution: $\nabla^2\Phi(E)=2\pi\rho_{\mathrm{OBC}}(E)$, where $\rho_{\mathrm{OBC}}$ is the DOS defined in Eq.~(\ref{DOS1}).
The loss of local harmonicity occurs at $|\beta_p|=|\beta_{p+1}|$, where the division into inner and outer root sets changes~\cite{Yuncheng2024PRB}.

\subsubsection{Summary and Scope of Validity}

Table~\ref{tab:gbz_criteria} summarizes the four equivalent criteria for the continuous OBC spectrum of 1D single-band models.
Each describes the same condition $|\beta_p|=|\beta_{p+1}|$ through a different mathematical object.
However, in multi-band systems, mode coupling and boundary constraints can invalidate the conventional root selection, as demonstrated in Sec.~\ref{sec:Limitations and Challenges of the Spectral Approach}.

\section{Spectral Approach III: Limitations and Challenges}
\label{sec:Limitations and Challenges of the Spectral Approach}

The spectral approach determines the continuous OBC spectrum of 1D single-band systems through the GBZ condition.
However, when it extends to the multi-band models, the central question is what is the corresponding GBZ condition for variant models. 

\subsection{Conventional and Anomalous Multi-Band GBZ Conditions}

\begin{figure}[t]
	\begin{center}
		\includegraphics[width=1\linewidth]{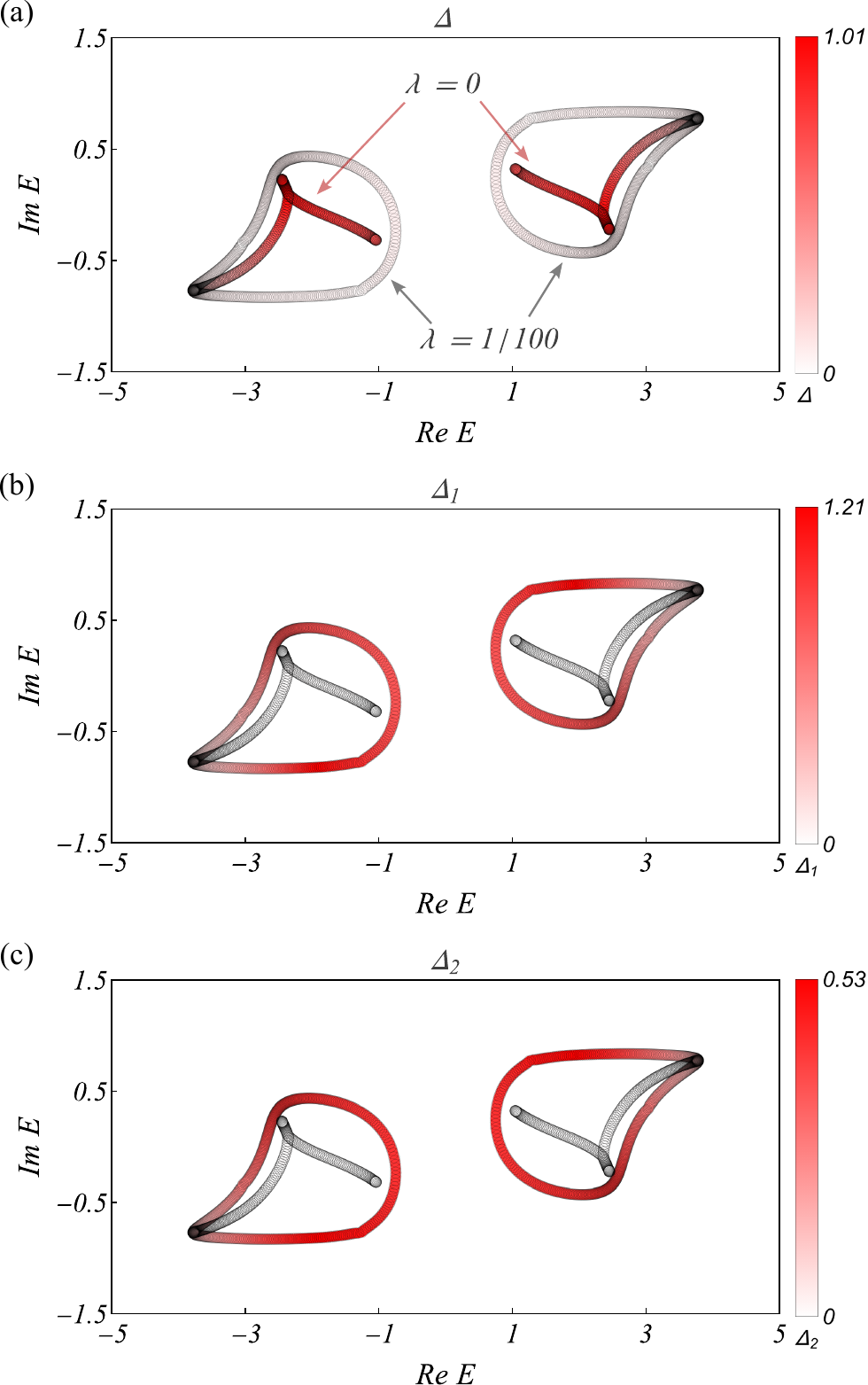}
		\par\end{center}
	\protect\caption{Root-modulus diagnostics for the non-Hermitian Rice--Mele model with time-reversal symmetry of transposition type ($\mathrm{TRS}^{\dagger}$) and its symmetry-broken counterpart. The (a), (b), and (c) show $\Delta(E)=|\beta_{np+1}|-|\beta_{np}|$, $\Delta_1(E)=|\beta_{np+2}|-|\beta_{np+1}|$, and $\Delta_2(E)=|\beta_{np}|-|\beta_{np-1}|$, respectively. Black open circles mark OBC eigenvalues, and red intensity represents the corresponding root-modulus difference. The inner spectra show $\lambda=0$, where the anomalous differences $\Delta_1$ and $\Delta_2$ are very small, while the conventional difference $\Delta$ remains finite. The outer spectra show $\lambda=1/100$, for which breaking $\mathrm{TRS}^{\dagger}$ lifts the degeneracy and shifts the root selection toward the conventional condition $\Delta=0$. Each panel uses an independent linear color scale; lighter colors indicate smaller differences, and absolute magnitudes are given by the color bars. Parameters are $t_1=\eta=2$, $t_2=\mu=\gamma=1$, and $L=200$. All computations use 30-digit numerical precision.
	}
	\label{F_TRS}
\end{figure}

\begin{figure}[t]
	\begin{center}
		\includegraphics[width=1\linewidth]{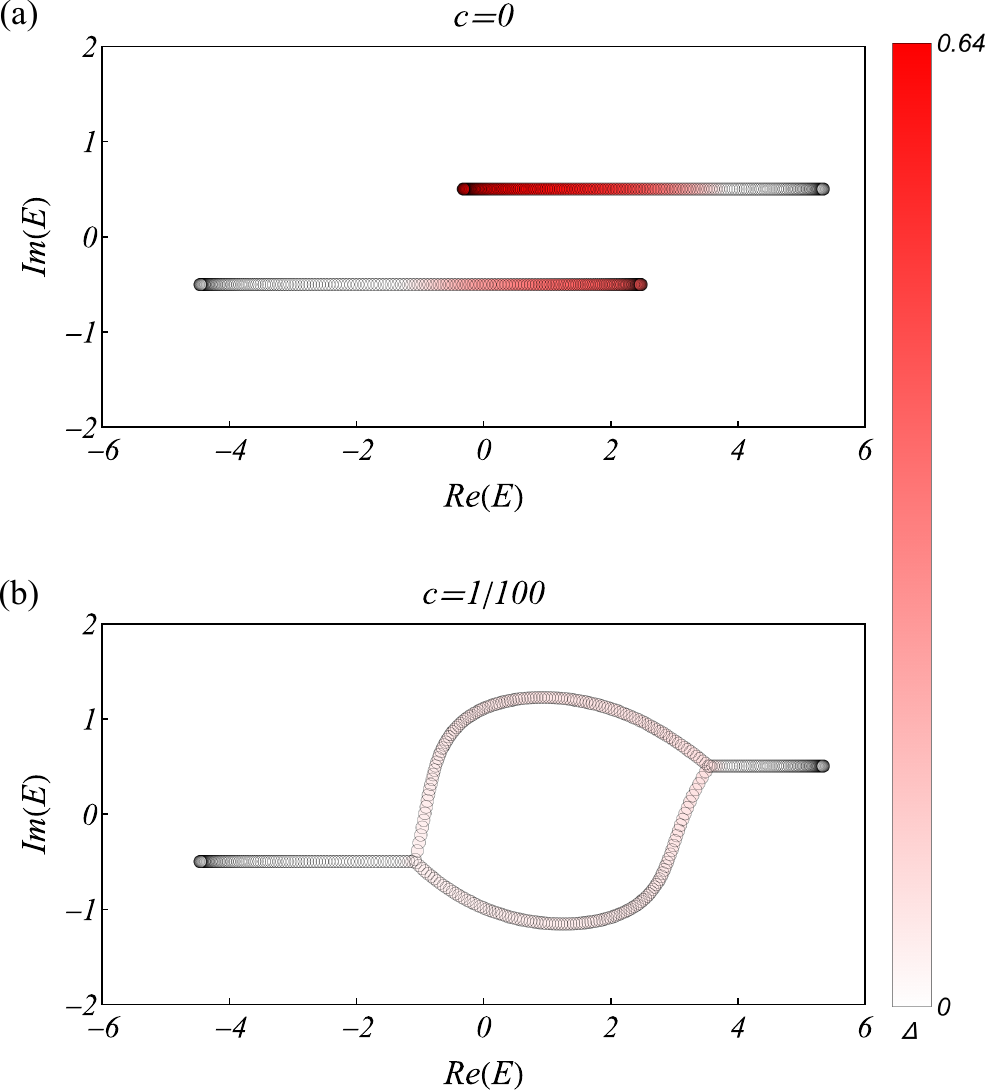}
		\par\end{center}
	\protect\caption{Deviation from the conventional 1D GBZ condition for the critical NHSE. Black open circles mark OBC eigenvalues, and red intensity represents $\Delta(E)$. In (a), $c=0$ and the decoupled OBC spectra lie on $\operatorname{Im}E=\gamma_t$ and $\operatorname{Im}E=\gamma_w$. In (b), a weak coupling $c=1/100$ reconstructs the separated spectra into a complex contour. The lighter colors indicate smaller $\Delta(E)$. Parameters are $t_0=5/2$, $t_1=1$, $t_{-1}=2$, $w_0=-1$, $w_1=3$, $w_{-1}=1$, $\gamma_t=0.5$, $\gamma_w=-0.5$, and $L=200$. All computations use 30-digit numerical precision.
	}
	\label{F_criticalSE}
\end{figure}

For a multi-band model with $n$ internal degrees of freedom, the ChE reads
\begin{equation}
	f(E,\beta):=\det[E\mathbbm{I}_n-H(\beta)]=0.
\end{equation}
With the  full-rank condition, the argument in Sec.~\ref{sec:Spectral Approach} can give the conventional condition
\begin{equation}\label{conventional_GBZ}
	|\beta_{np}(E)|=|\beta_{np+1}(E)|.
\end{equation}
However, some multi-band models can violate this condition~\cite{Kawabata2020PRBa,Yifei2020PRL}.
We refer to such an unconventional condition as an anomalous GBZ condition.

As a concrete example, consider a four-band extension of the non-Hermitian Rice--Mele model~\cite{Kawabata2020PRBa,Yifei2020PRL}.
We start with the spinless Hamiltonian
\begin{equation}
	H_{\mathrm{RM}}(k)=(t_1+t_2\cos k)\sigma_x+t_2\sin k\,\sigma_y+\mu\sigma_z,
\end{equation}
and introduce spin through
\begin{equation}\label{Rice-Mele_Hk}
	H_{\lambda}(k)=H_{\mathrm{RM}}(k)s_0+\eta\sin k\,\sigma_zs_z+i\gamma\sigma_zs_0+\lambda\sigma_0s_x.
\end{equation}
Here $\boldsymbol{\sigma}$ and $\boldsymbol{s}$ are Pauli matrices acting on the sublattice and spin spaces, respectively, and $\sigma_0$ and $s_0$ are identity matrices in those spaces.
At $\lambda=0$, Eq.~\eqref{Rice-Mele_Hk} satisfies
\begin{equation}
	U_T H_{\lambda=0}^T(k)U_T^{-1}=H_{\lambda=0}(-k),
\end{equation}
where $U_T=i\sigma_0s_y$ and $U_TU_T^*=-\mathbbm{I}_4$.
Thus, the model has symplectic $\mathrm{TRS}^{\dagger}$ and a degenerate OBC spectrum~\cite{Yifei2020PRL}.
The perturbation transforms as
\begin{equation}
	U_T(\lambda\sigma_0s_x)^TU_T^{-1}=-(\lambda\sigma_0s_x),
\end{equation}
so any nonzero $\lambda$ breaks this $\mathrm{TRS}^{\dagger}$ symmetry and generically lifts the degeneracy.
To examine root selection, we write the non-Bloch Hamiltonian as
\begin{align}\label{Rice-Mele_Hb}
	H_{\lambda}(\beta)
	&=\left[t_1+\frac{t_2}{2}(\beta+\beta^{-1})\right]\sigma_xs_0
	+\frac{t_2}{2i}(\beta-\beta^{-1})\sigma_ys_0\notag\\
	&\quad +(\mu+i\gamma)\sigma_zs_0
	+\frac{\eta}{2i}(\beta-\beta^{-1})\sigma_zs_z
	+\lambda\sigma_0s_x,
\end{align}
where $t_1=2$, $t_2=1$, $\mu=1$, $\gamma=1$, $\eta=2$.
This Hamiltonian has  $n=4$ and $p=q=1$.

For a general multi-band model, we define the neighboring root gaps
\begin{align}
	\Delta_1(E)&=|\beta_{np+2}(E)|-|\beta_{np+1}(E)|,\notag\\
	\Delta_2(E)&=|\beta_{np}(E)|-|\beta_{np-1}(E)|.
\end{align}
The conventional GBZ condition gives $\Delta=0$, whereas $\Delta_1=0$ and $\Delta_2=0$ describe the anomalous GBZ conditions considered here.
Fig.~\ref{F_TRS} compares these three root gaps on the OBC spectrum.

For each OBC eigenvalue $E$, we evaluate $\Delta_1$, $\Delta_2$, and $\Delta$.
Red intensity represents the corresponding gap, with lighter colors indicating smaller values within each panel.
At $\lambda=0$, it shows that $\Delta_1$ and $\Delta_2$ are strongly suppressed, whereas $\Delta$ remains finite.
At $\lambda=10^{-2}$, it shows a small $\Delta$ and large $\Delta_1$ and $\Delta_2$.

This simultaneous change suggests that spectral degeneracy may explain the anomalous condition.
The critical NHSE model of Refs.~\cite{Linhu2020NC,Zhesen2020PRL} appears to support this interpretation.
These examples suggest an association between spectral degeneracy and anomalous GBZ conditions.

\subsection{Spectral Degeneracy Is Not Necessary}

\begin{figure}[b]
	\begin{center}
		\includegraphics[width=0.91\linewidth]{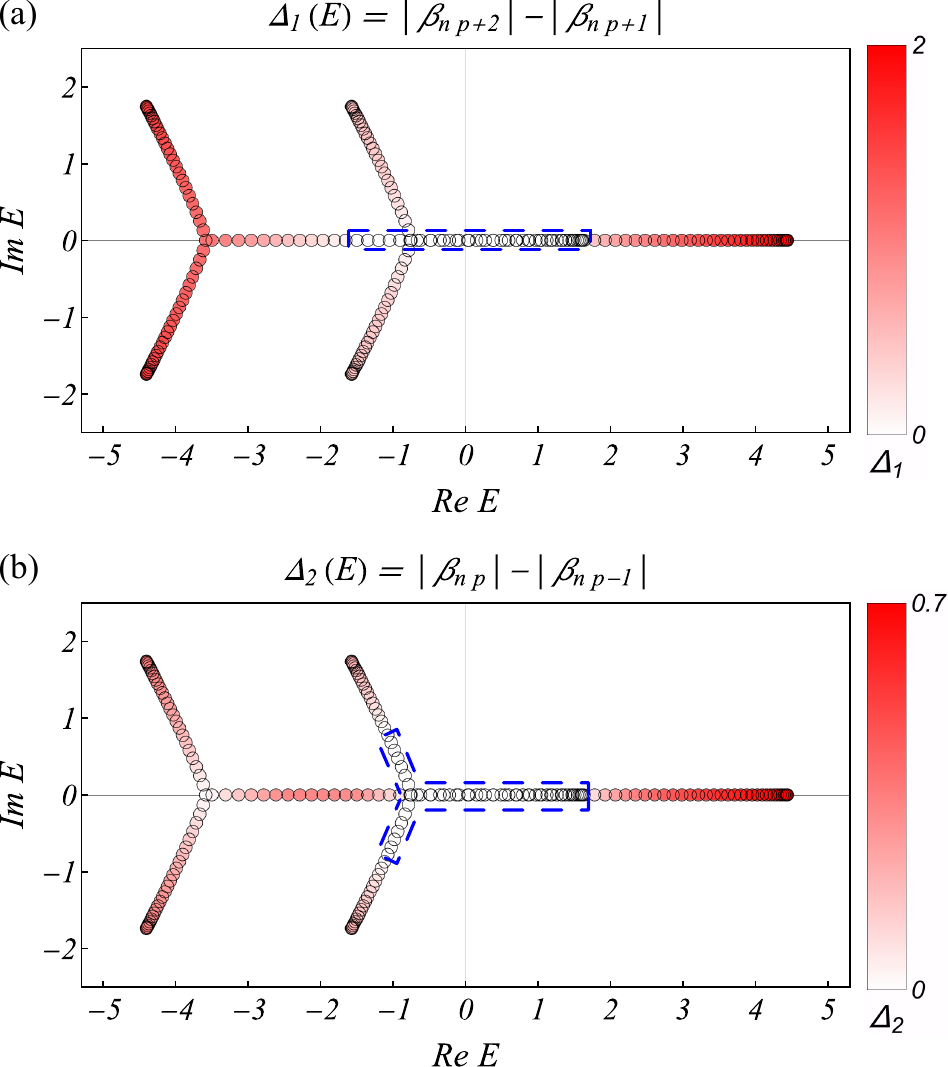}
		\par\end{center}
	\protect\caption{Anomalous root pairings for Unit Cell A. (a),(b) The OBC spectrum in Fig.~\ref{F_twoG2}(a), with color intensity representing $\Delta_1(E)=|\beta_{np+2}|-|\beta_{np+1}|$ and $\Delta_2(E)=|\beta_{np}|-|\beta_{np-1}|$, respectively. Blue dashed boxes highlight intervals with a nonzero conventional gap $\Delta$ and a zero gap $\Delta_1$ or $\Delta_2$. All computations use 16-digit numerical precision.
	}
	\label{F_anomaGBZ}
\end{figure}

\begin{figure*}[t]
	\begin{center}
		\includegraphics[width=1\linewidth]{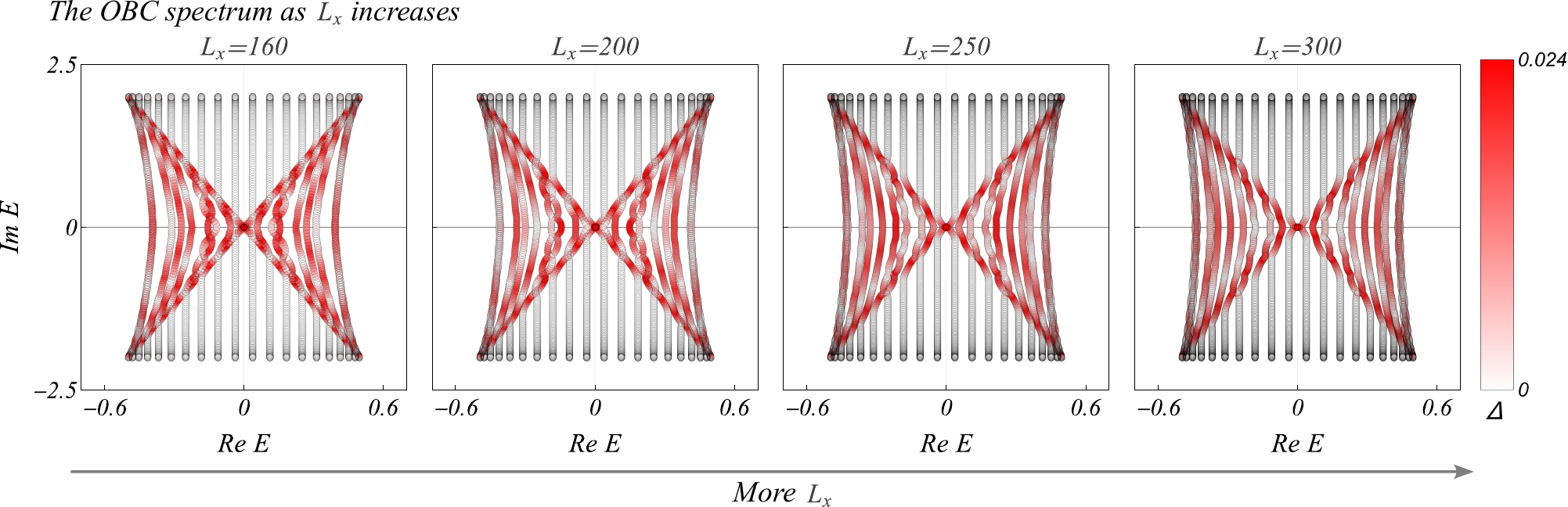}
		\par\end{center}
	\protect\caption{Size dependence of the OBC spectrum at fixed $n_y=20$. Black circles show the spectra for $L_x=160,200,250$, and $300$, and red intensity represents the 1D GBZ deviation $\Delta$. Parameters are $t_x=t_{-x}=i$ and $t_y=t_{-y}=1/4$. All computations use 16-digit numerical precision.
	}
	\label{F_increase_Lx}
\end{figure*}

To test whether spectral degeneracy is necessary, we separate the two OBC spectra while keeping their hopping asymmetries fixed.
We add unequal imaginary onsite potentials to the two diagonal sectors of the critical NHSE model of Refs.~\cite{Linhu2020NC,Zhesen2020PRL}. 
The non-Bloch Hamiltonian is given by 
\begin{equation}
	\label{critical_H1}
	\begin{aligned}
		H_c(\beta)
		&=\begin{pmatrix}
			h_t(\beta)+i\gamma_t & c\\
			c & h_w(\beta)+i\gamma_w
	\end{pmatrix},\\
	h_t(\beta)&=t_0+t_{-1}\beta^{-1}+t_1\beta,\\
	h_w(\beta)&=w_0+w_{-1}\beta^{-1}+w_1\beta,
	\end{aligned}
\end{equation}
where we choose $t_0=2.5$, $t_1=1$, $t_{-1}=2$, $w_0=-1$, $w_1=3$, $w_{-1}=1$, $\gamma_t=0.5$ and $\gamma_w=-0.5$.

At $c=0$, the ChE factorizes as
\begin{equation}\label{ChE_fc0}
	f_{c=0}(E,\beta)
	=[E-h_t(\beta)-i\gamma_t]
	[E-h_w(\beta)-i\gamma_w].
\end{equation}
The imaginary shifts leave the sub-GBZ radii unchanged at $\sqrt{2}$ and $1/\sqrt{3}$.
Each decoupled sector therefore retains its equal-modulus root pair. 
The upper and lower energy bands in Fig.~\ref{F_criticalSE} (a) satisfy the anomalous GBZ condition $\Delta_1=0$ and $\Delta_2=0$ respectively. 
However, not all OBC eigenvalues satisfy $\Delta=0$.
The anomalous intervals persist even though the two OBC spectra lie on distinct lines, $\operatorname{Im}E=\gamma_t$ and $\operatorname{Im}E=\gamma_w$.

For $c\neq0$, the ChE becomes
\begin{equation}\label{ChE_fc}
	f_c(E,\beta)=[E-h_t(\beta)-i\gamma_t][E-h_w(\beta)-i\gamma_w]-c^2,
\end{equation}
which is generically irreducible.
For $L=200$ and $c=1/100$, Fig.~\ref{F_criticalSE} (b) shows the reconstruction of the two horizontal spectra into a complex contour.
The conventional gap $\Delta$ is strongly suppressed along most of this contour, although a finite-size residual remains.
Thus, spectral degeneracy is unnecessary.

Separating the two spectra rules out degeneracy as a necessary condition for the anomalous pairing.
However, the weak coupling also changes the factorized ChE in Eq.~\eqref{ChE_fc0} into the irreducible ChE in Eq.~\eqref{ChE_fc}.
ChE factorization therefore remains a second possible explanation.

\subsection{Factorization of the ChE Is Not Sufficient}

The spectra in Figs.~\ref{F_twoG2}(a) and~\ref{F_twoG2}(b) test whether ChE factorization determines anomalous root selection.
They correspond to the same bulk two-chain lattice with two choices of unit cell.
Repeating intact Unit Cells A and B gives rectangular and parallelogram OBC geometries, respectively.
Although their off-diagonal matrix elements differ, the two non-Bloch Hamiltonians share the ChE~\eqref{ChE_equal}.
Thus, they have identical factorization and identical characteristic-root sets at fixed $E$.

Nevertheless, their OBC spectra and root selections differ.
For the boundary terminations considered here, only Unit Cell A exhibits the anomalous pairing as shown in Fig.~\ref{F_twoG2}. 
And Fig.~\ref{F_anomaGBZ} resolves the anomalous intervals for Unit Cell A.
Within the blue dashed boxes, the conventional gap $\Delta$ remains nonzero, whereas $\Delta_1$ or $\Delta_2$ is $0$.
These results show that ChE factorization alone cannot distinguish the two boundary problems.

\section{Summary and Discussion}\label{sec:Conclusion}

In this first paper of the series, we have reviewed the spectral approach to GBZ theory for 1D non-Hermitian systems.
Starting from the OBC eigenvalue problem, we constructed the BME and obtained exact finite-size quantization conditions for representative models.
For single-band systems, we derived the thermodynamic GBZ condition, the associated momentum quantization, and the OBC DOS.
We also discussed equivalent formulations based on characteristic roots, amoebas, winding numbers, and spectral potentials.
Finally, multi-band examples revealed the limitations of the conventional GBZ condition and the need for a general rule for selecting the appropriate root pairing.
These results lead to three open questions.

\textit{(i) Selection principle.}
The spectral approach and existing 1D GBZ theories provide several equivalent formulations for single-band systems, as summarized in Table~\ref{tab:gbz_criteria}.
However, a basic question remains unanswered in general: given a Bloch Hamiltonian and an OBC termination, which GBZ condition determines the spectrum?
This question becomes particularly important in the crossover from 1D to 2D.
Fig.~\ref{Fig_1Dto2D} shows deviations from the conventional GBZ condition as the system width increases.
At a fixed width $n_y=20$, Fig.~\ref{F_increase_Lx} shows that these deviations persist as $L_x$ increases, even at $L_x=300$.
Therefore, we believe that existing 1D GBZ theories remain incomplete, particularly in their formulation of the GBZ condition.

\textit{(ii) Finite-size characterization.}
Even if the conventional GBZ condition~\eqref{conventional_GBZ} applies to the system in Fig.~\ref{F_increase_Lx}, its finite-size convergence remains to be understood.
At fixed width $n_y=20$, the deviation $\Delta(E)$ from this condition must then vanish as $L_x\to\infty$, so the red color should gradually fade as $L_x$ increases.
This raises a practical question: how large must $L_x$ be for $\Delta(E)$ to become negligibly small and the spectrum to approach the GBZ prediction?
Existing 1D GBZ theories specify the thermodynamic condition but do not provide a general estimate of this convergence scale.

\textit{(iii) Numerical verification.}
Even if a theory estimates the convergence scale $L_c$ needed to approach the thermodynamic spectrum, can numerical calculations reliably reach this size?
For non-Hermitian Hamiltonians, small round-off errors can produce substantial errors in the computed eigenvalues.
How can these errors be controlled as the system size increases?
If $L_c$ lies far beyond the reach of reliable diagonalization, how can we verify the GBZ theory in the thermodynamic limit?

These questions motivate the wavefunction approach, which goes beyond the spectral description and forms the core of this series.
In Paper~II, we will introduce its central tool, \textit{coefficient flow}, which describes the spatial evolution of the relative weights of exponential-wave components.
In Paper~III, we will use coefficient flow to formulate a conjecture for a local theory and test it numerically.
This theory will help estimate the critical size $L_c$ needed for finite-size spectra to approach their thermodynamic limit.
In Paper~IV, we will introduce a new theoretical tool for testing thermodynamic GBZ predictions without directly diagonalizing the OBC Hamiltonian.
This tool will address the third question when the required system size exceeds the reach of reliable diagonalization.

\appendix

\bibliographystyle{apsrev4-2}
\bibliography{refs}

@article{bloch1929quantenmechanik,
  title     = {{\"U}ber die quantenmechanik der elektronen in kristallgittern},
  author    = {Bloch, Felix},
  journal   = {Zeitschrift f{\"u}r physik},
  volume    = {52},
  number    = {7},
  pages     = {555--600},
  year      = {1929},
  publisher = {Springer}
}

@book{kittel1976introduction,
  title     = {Introduction to solid state physics},
  author    = {Kittel, Charles and McEuen, Paul},
  volume    = {8},
  year      = {1976},
  publisher = {wiley New York}
}

@book{ashcroft1976solid,
  title     = {Solid State Physics},
  author    = {Ashcroft, Neil W. and Mermin, N. David},
  year      = {1976},
  publisher = {Holt, Rinehart and Winston},
  address   = {New York}
}

@article{ZhongWang2018PRL,
  title     = {Edge States and Topological Invariants of Non-Hermitian Systems},
  author    = {Yao, Shunyu and Wang, Zhong},
  journal   = {Phys. Rev. Lett.},
  volume    = {121},
  issue     = {8},
  pages     = {086803},
  numpages  = {8},
  year      = {2018},
  month     = {Aug},
  publisher = {American Physical Society},
  doi       = {10.1103/PhysRevLett.121.086803},
  url       = {https://link.aps.org/doi/10.1103/PhysRevLett.121.086803}
}

@article{Shunyu2018PRL,
  title     = {Non-Hermitian Chern Bands},
  author    = {Yao, Shunyu and Song, Fei and Wang, Zhong},
  journal   = {Phys. Rev. Lett.},
  volume    = {121},
  issue     = {13},
  pages     = {136802},
  numpages  = {8},
  year      = {2018},
  month     = {Sep},
  publisher = {American Physical Society},
  doi       = {10.1103/PhysRevLett.121.136802},
  url       = {https://link.aps.org/doi/10.1103/PhysRevLett.121.136802}
}

@article{Gong2018PRX,
  title     = {Topological Phases of Non-Hermitian Systems},
  author    = {Gong, Zongping and Ashida, Yuto and Kawabata, Kohei and Takasan, Kazuaki and Higashikawa, Sho and Ueda, Masahito},
  journal   = {Phys. Rev. X},
  volume    = {8},
  issue     = {3},
  pages     = {031079},
  numpages  = {33},
  year      = {2018},
  month     = {Sep},
  publisher = {American Physical Society},
  doi       = {10.1103/PhysRevX.8.031079},
  url       = {https://link.aps.org/doi/10.1103/PhysRevX.8.031079}
}

@article{Kunst2018PRL,
  title     = {Biorthogonal Bulk-Boundary Correspondence in Non-Hermitian Systems},
  author    = {Kunst, Flore K. and Edvardsson, Elisabet and Budich, Jan Carl and Bergholtz, Emil J.},
  journal   = {Phys. Rev. Lett.},
  volume    = {121},
  issue     = {2},
  pages     = {026808},
  numpages  = {6},
  year      = {2018},
  month     = {Jul},
  publisher = {American Physical Society},
  doi       = {10.1103/PhysRevLett.121.026808},
  url       = {https://link.aps.org/doi/10.1103/PhysRevLett.121.026808}
}

@article{Martinez2018PRB,
  title     = {Non-Hermitian robust edge states in one dimension: Anomalous localization and eigenspace condensation at exceptional points},
  author    = {Martinez Alvarez, V. M. and Barrios Vargas, J. E. and Foa Torres, L. E. F.},
  journal   = {Phys. Rev. B},
  volume    = {97},
  issue     = {12},
  pages     = {121401(R)},
  numpages  = {6},
  year      = {2018},
  month     = {Mar},
  publisher = {American Physical Society},
  doi       = {10.1103/PhysRevB.97.121401},
  url       = {https://link.aps.org/doi/10.1103/PhysRevB.97.121401}
}

@article{Deng2019PRB,
  title     = {Non-Bloch topological invariants in a non-Hermitian domain wall system},
  author    = {Deng, Tian-Shu and Yi, Wei},
  journal   = {Phys. Rev. B},
  volume    = {100},
  issue     = {3},
  pages     = {035102},
  numpages  = {8},
  year      = {2019},
  month     = {Jul},
  publisher = {American Physical Society},
  doi       = {10.1103/PhysRevB.100.035102},
  url       = {https://link.aps.org/doi/10.1103/PhysRevB.100.035102}
}

@article{Longhi2019PRR,
  title     = {Probing non-Hermitian skin effect and non-Bloch phase transitions},
  author    = {Longhi, Stefano},
  journal   = {Phys. Rev. Res.},
  volume    = {1},
  issue     = {2},
  pages     = {023013},
  numpages  = {13},
  year      = {2019},
  month     = {Sep},
  publisher = {American Physical Society},
  doi       = {10.1103/PhysRevResearch.1.023013},
  url       = {https://link.aps.org/doi/10.1103/PhysRevResearch.1.023013}
}

@article{Kawabata2019PRX,
  title     = {Symmetry and Topology in Non-Hermitian Physics},
  author    = {Kawabata, Kohei and Shiozaki, Ken and Ueda, Masahito and Sato, Masatoshi},
  journal   = {Phys. Rev. X},
  volume    = {9},
  issue     = {4},
  pages     = {041015},
  numpages  = {52},
  year      = {2019},
  month     = {Oct},
  publisher = {American Physical Society},
  doi       = {10.1103/PhysRevX.9.041015},
  url       = {https://link.aps.org/doi/10.1103/PhysRevX.9.041015}
}

@article{Fei2019PRL,
  title     = {Non-Hermitian Skin Effect and Chiral Damping in Open Quantum Systems},
  author    = {Song, Fei and Yao, Shunyu and Wang, Zhong},
  journal   = {Phys. Rev. Lett.},
  volume    = {123},
  issue     = {17},
  pages     = {170401},
  numpages  = {8},
  year      = {2019},
  month     = {Oct},
  publisher = {American Physical Society},
  doi       = {10.1103/PhysRevLett.123.170401},
  url       = {https://link.aps.org/doi/10.1103/PhysRevLett.123.170401}
}

@article{Murakami2019PRL,
  title     = {Non-Bloch Band Theory of Non-Hermitian Systems},
  author    = {Yokomizo, Kazuki and Murakami, Shuichi},
  journal   = {Phys. Rev. Lett.},
  volume    = {123},
  issue     = {6},
  pages     = {066404},
  numpages  = {6},
  year      = {2019},
  month     = {Aug},
  publisher = {American Physical Society},
  doi       = {10.1103/PhysRevLett.123.066404},
  url       = {https://link.aps.org/doi/10.1103/PhysRevLett.123.066404}
}

@article{SongFei2019PRL,
  title     = {Non-Hermitian Topological Invariants in Real Space},
  author    = {Song, Fei and Yao, Shunyu and Wang, Zhong},
  journal   = {Phys. Rev. Lett.},
  volume    = {123},
  issue     = {24},
  pages     = {246801},
  numpages  = {8},
  year      = {2019},
  month     = {Dec},
  publisher = {American Physical Society},
  doi       = {10.1103/PhysRevLett.123.246801},
  url       = {https://link.aps.org/doi/10.1103/PhysRevLett.123.246801}
}

@article{ChingHua2019PRB,
  title     = {Anatomy of skin modes and topology in non-Hermitian systems},
  author    = {Lee, Ching Hua and Thomale, Ronny},
  journal   = {Phys. Rev. B},
  volume    = {99},
  issue     = {20},
  pages     = {201103(R)},
  numpages  = {5},
  year      = {2019},
  month     = {May},
  publisher = {American Physical Society},
  doi       = {10.1103/PhysRevB.99.201103},
  url       = {https://link.aps.org/doi/10.1103/PhysRevB.99.201103}
}

@article{Kai2020PRL,
  title     = {Correspondence between Winding Numbers and Skin Modes in Non-Hermitian Systems},
  author    = {Zhang, Kai and Yang, Zhesen and Fang, Chen},
  journal   = {Phys. Rev. Lett.},
  volume    = {125},
  issue     = {12},
  pages     = {126402},
  numpages  = {6},
  year      = {2020},
  month     = {Sep},
  publisher = {American Physical Society},
  doi       = {10.1103/PhysRevLett.125.126402},
  url       = {https://link.aps.org/doi/10.1103/PhysRevLett.125.126402}
}

@article{Zhesen2020PRL,
  title     = {Non-Hermitian Bulk-Boundary Correspondence and Auxiliary Generalized Brillouin Zone Theory},
  author    = {Yang, Zhesen and Zhang, Kai and Fang, Chen and Hu, Jiangping},
  journal   = {Phys. Rev. Lett.},
  volume    = {125},
  issue     = {22},
  pages     = {226402},
  numpages  = {6},
  year      = {2020},
  month     = {Nov},
  publisher = {American Physical Society},
  doi       = {10.1103/PhysRevLett.125.226402},
  url       = {https://link.aps.org/doi/10.1103/PhysRevLett.125.226402}
}

@article{Kawabata2020PRBa,
  title     = {Non-Bloch band theory of non-Hermitian Hamiltonians in the symplectic class},
  author    = {Kawabata, Kohei and Okuma, Nobuyuki and Sato, Masatoshi},
  journal   = {Phys. Rev. B},
  volume    = {101},
  issue     = {19},
  pages     = {195147},
  numpages  = {12},
  year      = {2020},
  month     = {May},
  publisher = {American Physical Society},
  doi       = {10.1103/PhysRevB.101.195147},
  url       = {https://link.aps.org/doi/10.1103/PhysRevB.101.195147}
}

@article{Yifei2020PRL,
  title     = {Non-Hermitian Skin Modes Induced by On-Site Dissipations and Chiral Tunneling Effect},
  author    = {Yi, Yifei and Yang, Zhesen},
  journal   = {Phys. Rev. Lett.},
  volume    = {125},
  issue     = {18},
  pages     = {186802},
  numpages  = {7},
  year      = {2020},
  month     = {Oct},
  publisher = {American Physical Society},
  doi       = {10.1103/PhysRevLett.125.186802},
  url       = {https://link.aps.org/doi/10.1103/PhysRevLett.125.186802}
}

@article{Okuma2020PRL,
  title     = {Topological Origin of Non-Hermitian Skin Effects},
  author    = {Okuma, Nobuyuki and Kawabata, Kohei and Shiozaki, Ken and Sato, Masatoshi},
  journal   = {Phys. Rev. Lett.},
  volume    = {124},
  issue     = {8},
  pages     = {086801},
  numpages  = {7},
  year      = {2020},
  month     = {Feb},
  publisher = {American Physical Society},
  doi       = {10.1103/PhysRevLett.124.086801},
  url       = {https://link.aps.org/doi/10.1103/PhysRevLett.124.086801}
}

@article{Borgnia2020PRL,
  title     = {Non-Hermitian Boundary Modes and Topology},
  author    = {Borgnia, Dan S. and Kruchkov, Alex Jura and Slager, Robert-Jan},
  journal   = {Phys. Rev. Lett.},
  volume    = {124},
  issue     = {5},
  pages     = {056802},
  numpages  = {6},
  year      = {2020},
  month     = {Feb},
  publisher = {American Physical Society},
  doi       = {10.1103/PhysRevLett.124.056802},
  url       = {https://link.aps.org/doi/10.1103/PhysRevLett.124.056802}
}

@article{Zirnstein2021PRL,
  title     = {Bulk-Boundary Correspondence for Non-Hermitian Hamiltonians via Green Functions},
  author    = {Zirnstein, Heinrich-Gregor and Refael, Gil and Rosenow, Bernd},
  journal   = {Phys. Rev. Lett.},
  volume    = {126},
  issue     = {21},
  pages     = {216407},
  numpages  = {7},
  year      = {2021},
  month     = {May},
  publisher = {American Physical Society},
  doi       = {10.1103/PhysRevLett.126.216407},
  url       = {https://link.aps.org/doi/10.1103/PhysRevLett.126.216407}
}

@article{Liu2024PRL,
  title     = {Localization of Chiral Edge States by the Non-Hermitian Skin Effect},
  author    = {Liu, Gui-Geng and Mandal, Subhaskar and Zhou, Peiheng and Xi, Xiang and Banerjee, Rimi and Hu, Yuan-Hang and Wei, Minggui and Wang, Maoren and Wang, Qiang and Gao, Zhen and Chen, Hongsheng and Yang, Yihao and Chong, Yidong and Zhang, Baile},
  journal   = {Phys. Rev. Lett.},
  volume    = {132},
  issue     = {11},
  pages     = {113802},
  numpages  = {6},
  year      = {2024},
  month     = {Mar},
  publisher = {American Physical Society},
  doi       = {10.1103/PhysRevLett.132.113802},
  url       = {https://link.aps.org/doi/10.1103/PhysRevLett.132.113802}
}

@article{Kawabata2020PRBb,
  title     = {Higher-order non-Hermitian skin effect},
  author    = {Kawabata, Kohei and Sato, Masatoshi and Shiozaki, Ken},
  journal   = {Phys. Rev. B},
  volume    = {102},
  issue     = {20},
  pages     = {205118},
  numpages  = {16},
  year      = {2020},
  month     = {Nov},
  publisher = {American Physical Society},
  doi       = {10.1103/PhysRevB.102.205118},
  url       = {https://link.aps.org/doi/10.1103/PhysRevB.102.205118}
}

@article{Kai2022NC,
  title   = {Universal non-{Hermitian} skin effect in two and higher dimensions},
  volume  = {13},
  issn    = {2041-1723},
  url     = {https://doi.org/10.1038/s41467-022-30161-6},
  doi     = {10.1038/s41467-022-30161-6},
  number  = {1},
  journal = {Nature Communications},
  author  = {Zhang, Kai and Yang, Zhesen and Fang, Chen},
  month   = may,
  year    = {2022},
  pages   = {2496}
}

@article{Xue2022PRL,
  title     = {Non-Hermitian Edge Burst},
  author    = {Xue, Wen-Tan and Hu, Yu-Min and Song, Fei and Wang, Zhong},
  journal   = {Phys. Rev. Lett.},
  volume    = {128},
  issue     = {12},
  pages     = {120401},
  numpages  = {6},
  year      = {2022},
  month     = {Mar},
  publisher = {American Physical Society},
  doi       = {10.1103/PhysRevLett.128.120401},
  url       = {https://link.aps.org/doi/10.1103/PhysRevLett.128.120401}
}

@article{Hui2023PRL,
  title     = {Dimensional Transmutation from Non-Hermiticity},
  author    = {Jiang, Hui and Lee, Ching Hua},
  journal   = {Phys. Rev. Lett.},
  volume    = {131},
  issue     = {7},
  pages     = {076401},
  numpages  = {12},
  year      = {2023},
  month     = {Aug},
  publisher = {American Physical Society},
  doi       = {10.1103/PhysRevLett.131.076401},
  url       = {https://link.aps.org/doi/10.1103/PhysRevLett.131.076401}
}

@article{Yokomizo2023PRB,
  title     = {Non-Bloch bands in two-dimensional non-Hermitian systems},
  author    = {Yokomizo, Kazuki and Murakami, Shuichi},
  journal   = {Phys. Rev. B},
  volume    = {107},
  issue     = {19},
  pages     = {195112},
  numpages  = {10},
  year      = {2023},
  month     = {May},
  publisher = {American Physical Society},
  doi       = {10.1103/PhysRevB.107.195112},
  url       = {https://link.aps.org/doi/10.1103/PhysRevB.107.195112}
}

@article{ZhangKai2023PRL,
  title     = {Dynamical Degeneracy Splitting and Directional Invisibility in Non-Hermitian Systems},
  author    = {Zhang, Kai and Fang, Chen and Yang, Zhesen},
  journal   = {Phys. Rev. Lett.},
  volume    = {131},
  issue     = {3},
  pages     = {036402},
  numpages  = {6},
  year      = {2023},
  month     = {Jul},
  publisher = {American Physical Society},
  doi       = {10.1103/PhysRevLett.131.036402},
  url       = {https://link.aps.org/doi/10.1103/PhysRevLett.131.036402}
}

@article{Kawabata2023PRX,
  title     = {Entanglement Phase Transition Induced by the Non-Hermitian Skin Effect},
  author    = {Kawabata, Kohei and Numasawa, Tokiro and Ryu, Shinsei},
  journal   = {Phys. Rev. X},
  volume    = {13},
  issue     = {2},
  pages     = {021007},
  numpages  = {26},
  year      = {2023},
  month     = {Apr},
  publisher = {American Physical Society},
  doi       = {10.1103/PhysRevX.13.021007},
  url       = {https://link.aps.org/doi/10.1103/PhysRevX.13.021007}
}

@article{Fang2023PRB,
  title     = {Point-gap bound states in non-Hermitian systems},
  author    = {Fang, Zixi and Fang, Chen and Zhang, Kai},
  journal   = {Phys. Rev. B},
  volume    = {108},
  issue     = {16},
  pages     = {165132},
  numpages  = {6},
  year      = {2023},
  month     = {Oct},
  publisher = {American Physical Society},
  doi       = {10.1103/PhysRevB.108.165132},
  url       = {https://link.aps.org/doi/10.1103/PhysRevB.108.165132}
}

@article{Yuncheng2024PRB,
  title     = {Graph morphology of non-Hermitian bands},
  author    = {Xiong, Yuncheng and Hu, Haiping},
  journal   = {Phys. Rev. B},
  volume    = {109},
  issue     = {10},
  pages     = {L100301},
  numpages  = {6},
  year      = {2024},
  month     = {Mar},
  publisher = {American Physical Society},
  doi       = {10.1103/PhysRevB.109.L100301},
  url       = {https://link.aps.org/doi/10.1103/PhysRevB.109.L100301}
}

@article{Hu2024PRL,
  title     = {Geometric Origin of Non-Bloch $\mathcal{P}\mathcal{T}$ Symmetry Breaking},
  author    = {Hu, Yu-Min and Wang, Hong-Yi and Wang, Zhong and Song, Fei},
  journal   = {Phys. Rev. Lett.},
  volume    = {132},
  issue     = {5},
  pages     = {050402},
  numpages  = {7},
  year      = {2024},
  month     = {Jan},
  publisher = {American Physical Society},
  doi       = {10.1103/PhysRevLett.132.050402},
  url       = {https://link.aps.org/doi/10.1103/PhysRevLett.132.050402}
}

@article{Hongyi2024PRX,
  title     = {Amoeba Formulation of Non-Bloch Band Theory in Arbitrary Dimensions},
  author    = {Wang, Hong-Yi and Song, Fei and Wang, Zhong},
  journal   = {Phys. Rev. X},
  volume    = {14},
  issue     = {2},
  pages     = {021011},
  numpages  = {21},
  year      = {2024},
  month     = {Apr},
  publisher = {American Physical Society},
  doi       = {10.1103/PhysRevX.14.021011},
  url       = {https://link.aps.org/doi/10.1103/PhysRevX.14.021011}
}

@article{Haiping2025SciB,
  title   = {Topological origin of non-Hermitian skin effect in higher dimensions and uniform spectra},
  journal = {Science Bulletin},
  volume  = {70},
  number  = {1},
  pages   = {51-57},
  year    = {2025},
  issn    = {2095-9273},
  doi     = {https://doi.org/10.1016/j.scib.2024.07.022},
  url     = {https://www.sciencedirect.com/science/article/pii/S2095927324005024},
  author  = {Haiping Hu}
}

@article{Kai2025PRX,
  title     = {Algebraic Non-Hermitian Skin Effect and Generalized Fermi Surface Formula in Arbitrary Dimensions},
  author    = {Zhang, Kai and Shu, Chang and Sun, Kai},
  journal   = {Phys. Rev. X},
  volume    = {15},
  issue     = {3},
  pages     = {031039},
  numpages  = {32},
  year      = {2025},
  month     = {Aug},
  publisher = {American Physical Society},
  doi       = {10.1103/cwwd-bclc},
  url       = {https://link.aps.org/doi/10.1103/cwwd-bclc}
}

@misc{Yuncheng2024arXiv,
  title         = {Non-Hermitian skin effect in arbitrary dimensions: non-Bloch band theory and classification},
  author        = {Yuncheng Xiong and Ze-Yu Xing and Haiping Hu},
  year          = {2024},
  eprint        = {2407.01296},
  archiveprefix = {arXiv},
  primaryclass  = {cond-mat.mes-hall},
  url           = {https://arxiv.org/abs/2407.01296}
}

@misc{Zeqi2023arXiv,
  title         = {Two-dimensional Asymptotic Generalized Brillouin Zone Theory},
  author        = {Zeqi Xu and Bo Pang and Kai Zhang and Zhesen Yang},
  year          = {2024},
  eprint        = {2311.16868},
  archiveprefix = {arXiv},
  primaryclass  = {cond-mat.mes-hall},
  url           = {https://arxiv.org/abs/2311.16868}
}

@misc{ChangShu2024arXiv,
  title         = {Ultra spectral sensitivity and non-local bi-impurity bound states from quasi-long-range non-hermitian skin modes},
  author        = {Chang Shu and Kai Zhang and Kai Sun},
  year          = {2024},
  eprint        = {2409.13623},
  archiveprefix = {arXiv},
  primaryclass  = {cond-mat.mes-hall},
  url           = {https://arxiv.org/abs/2409.13623}
}

@misc{Chenyang2025arXiv,
  title         = {General theory for geometry-dependent non-Hermitian bands},
  author        = {Chenyang Wang and Jinghui Pi and Qinxin Liu and Yaohua Li and Yong-Chun Liu},
  year          = {2025},
  eprint        = {2506.22743},
  archiveprefix = {arXiv},
  primaryclass  = {cond-mat.mes-hall},
  url           = {https://arxiv.org/abs/2506.22743}
}

@article{Linhu2020NC,
  author  = {Li, Linhu and Lee, Ching Hua and Mu, Sen and Gong, Jiangbin},
  title   = {Critical non-Hermitian skin effect},
  journal = {Nature Communications},
  volume  = {11},
  number  = {1},
  pages   = {5491},
  year    = {2020},
  doi     = {10.1038/s41467-020-18917-4},
  url     = {https://doi.org/10.1038/s41467-020-18917-4}
}

@article{Yifei2025PRBa,
  title     = {Critical non-Hermitian skin effect induced by boundary defects},
  author    = {Yi, Yifei},
  journal   = {Phys. Rev. B},
  volume    = {111},
  issue     = {14},
  pages     = {144307},
  numpages  = {8},
  year      = {2025},
  month     = {Apr},
  publisher = {American Physical Society},
  doi       = {10.1103/PhysRevB.111.144307},
  url       = {https://link.aps.org/doi/10.1103/PhysRevB.111.144307}
}

@article{Yifei2025PRBb,
  title     = {Anomalous scaling behavior of Green's function in critical skin effects},
  author    = {Yi, Yifei and Yang, Zhesen},
  journal   = {Phys. Rev. B},
  volume    = {112},
  issue     = {17},
  pages     = {174303},
  numpages  = {8},
  year      = {2025},
  month     = {Nov},
  publisher = {American Physical Society},
  doi       = {10.1103/d5zc-p1sk},
  url       = {https://link.aps.org/doi/10.1103/d5zc-p1sk}
}

@article{Wentan2021PRB,
  title     = {Simple formulas of directional amplification from non-Bloch band theory},
  author    = {Xue, Wen-Tan and Li, Ming-Rui and Hu, Yu-Min and Song, Fei and Wang, Zhong},
  journal   = {Phys. Rev. B},
  volume    = {103},
  issue     = {24},
  pages     = {L241408},
  numpages  = {6},
  year      = {2021},
  month     = {Jun},
  publisher = {American Physical Society},
  doi       = {10.1103/PhysRevB.103.L241408},
  url       = {https://link.aps.org/doi/10.1103/PhysRevB.103.L241408}
}

@article{Review1,
  title     = {Exceptional topology of non-Hermitian systems},
  author    = {Bergholtz, Emil J. and Budich, Jan Carl and Kunst, Flore K.},
  journal   = {Rev. Mod. Phys.},
  volume    = {93},
  issue     = {1},
  pages     = {015005},
  numpages  = {31},
  year      = {2021},
  month     = {Feb},
  publisher = {American Physical Society},
  doi       = {10.1103/RevModPhys.93.015005},
  url       = {https://link.aps.org/doi/10.1103/RevModPhys.93.015005}
}

@article{Review2,
  author  = {Ding, Kun and Fang, Chen and Ma, Guancong},
  title   = {Non-Hermitian topology and exceptional-point geometries},
  journal = {Nature Reviews Physics},
  volume  = {4},
  number  = {12},
  pages   = {745--760},
  year    = {2022},
  doi     = {10.1038/s42254-022-00516-5},
  url     = {https://doi.org/10.1038/s42254-022-00516-5}
}

@article{Review3,
  author    = {Xiujuan Zhang and Tian Zhang and Ming-Hui Lu and Yan-Feng Chen},
  title     = {A review on non-Hermitian skin effect},
  journal   = {Advances in Physics: X},
  volume    = {7},
  number    = {1},
  pages     = {2109431},
  year      = {2022},
  publisher = {Taylor \& Francis},
  doi       = {10.1080/23746149.2022.2109431},
  url       = {https://doi.org/10.1080/23746149.2022.2109431},
  eprint    = { https://doi.org/10.1080/23746149.2022.2109431}
}

@article{Review4,
  author  = {Lin, Rijia and Tai, Tommy and Li, Linhu and Lee, Ching Hua},
  title   = {Topological non-Hermitian skin effect},
  journal = {Frontiers of Physics},
  volume  = {18},
  number  = {5},
  pages   = {53605},
  year    = {2023},
  doi     = {10.1007/s11467-023-1309-z},
  url     = {https://doi.org/10.1007/s11467-023-1309-z}
}
\end{document}